\documentclass[smallextended]{svjour3}%
\pdfoutput=1
\usepackage{graphics,amsmath,amsfonts,amscd,revsymb,latexsym,
enumerate,multirow,epsfig}
\usepackage{amsmath}
\usepackage{amsfonts}
\usepackage{amssymb}
\usepackage{graphicx}
\usepackage{hyperref}
\usepackage{fullpage}
\usepackage{braket}
\usepackage{enumerate}

\spnewtheorem{definitionenv}{Definition}{\bf}{\rm}
\spnewtheorem{lemmaenv}[definitionenv]{Lemma}{\bf}{\rm}
\spnewtheorem{theoremenv}[definitionenv]{Theorem}{\bf}{\rm}
\spnewtheorem{corollaryenv}[definitionenv]{Corollary}{\bf}{\rm}
\spnewtheorem{propositionenv}[definitionenv]{Proposition}{\bf}{\rm}
\spnewtheorem{conjectureenv}[definitionenv]{Conjecture}{\bf}{\rm}
\spnewtheorem{exampleenv}{Example}{\bf}{\rm}
\spnewtheorem{app-lemmaenv}[section]{Lemma}{\bf}{\rm}

\newcommand{\bd}{\begin{definitionenv}}
\newcommand{\ed}{\end{definitionenv}}
\newcommand{\bl}{\begin{lemmaenv}}
\newcommand{\el}{\end{lemmaenv}}
\newcommand{\elp}{\hspace*{\fill} $\Box$
                 \end{lemmaenv}}
\newcommand{\bt}{\begin{theoremenv}}
\newcommand{\et}{\end{theoremenv}}
\newcommand{\etp}{\hspace*{\fill} $\Box$
                 \end{theoremenv}}
\newcommand{\bc}{\begin{corollaryenv}}
\newcommand{\ec}{\end{corollaryenv}}
\newcommand{\ecp}{\hspace*{\fill} $\Box$
                 \end{corollaryenv}}
\newcommand{\bcj}{\begin{conjecture}}
\newcommand{\ecj}{\end{conjecture}}

\newcommand{\be}{\begin{exampleenv}}
\newcommand{\ee}{\end{exampleenv}}
\newcommand{\eep}{\hspace*{\fill} $\Box$
                 \end{example}}
\newcommand{\bp}{\begin{proposition}}
\newcommand{\ep}{\end{proposition}}
\newcommand{\epp}{%\hspace*{\fill} $\Box$
                 \end{proposition}}

\newcommand{\wt}[1]{\mathrm{wt}({#1})}

\begin{document}

\title{Dualities and Identities for Entanglement-Assisted Quantum Codes%\thanks{Grants or other notes
%about the article that should go on the front page should be
%placed here. General acknowledgments should be placed at the end of the article.}
}
%\subtitle{Do you have a subtitle?\\ If so, write it here}

%\titlerunning{Short form of title}        % if too long for running head

\author{Ching-Yi Lai \and Todd A. Brun \and Mark M. Wilde %etc.
}

%\authorrunning{Short form of author list} % if too long for running head

\institute{Ching-Yi Lai and Todd A. Brun are with the
        Communication Sciences Institute, Electrical Engineering Department,
University of Southern California, Los Angeles, California, USA  90089.
              \email{laiching@usc.edu and tbrun@usc.edu}           %  \\
%             \emph{Present address:} of F. Author  %  if needed
\and
           Mark M. Wilde was with the School of Computer
Science, McGill University, Montreal, Quebec, Canada H3A 2A7 when this research
was conducted. He is now with the Hearne Institute for Theoretical Physics,
the Department of Physics and Astronomy, and the Center for Computation and
Technology at Louisiana State University, Baton Rouge, Louisiana 70803, USA
\email{mwilde@gmail.com}.
}

\date{Received: date / Accepted: date}
% The correct dates will be entered by the editor

\maketitle

\begin{abstract}
The dual of an entanglement-assisted quantum error-correcting (EAQEC) code
is the code resulting from exchanging the original code's information qubits with its ebits.
To introduce this notion, we show how
entanglement-assisted (EA) repetition codes and accumulator codes are dual to each other,
much like their classical counterparts, and we give an explicit, general quantum shift-register circuit that encodes both classes of codes.
We later show that our constructions are optimal, and this result completes our understanding of these dual classes of codes.
%For general EAQEC code, we studied linear programming bounds by exploiting these dualities and their corresponding MacWilliams identities.
We also establish the Gilbert-Varshamov bound and the Plotkin bound for EAQEC codes,
and we use these to examine the existence of some EAQEC codes.
%Combining these bounds allows us to formulate a table of upper and lower bounds on
%the minimum distance of any maximal-entanglement EAQEC code with length up to 15 channel qubits.
Finally, we provide upper bounds on
the block error probability when transmitting maximal-entanglement
EAQEC codes over the depolarizing channel, and
we derive variations of the hashing bound for EAQEC codes, which is a lower bound
on the maximum rate at which reliable communication over Pauli channels is possible with the use of pre-shared entanglement.
% \subclass{MSC code1 \and MSC code2 \and more}
\end{abstract}

\section{Introduction}

The existence of quantum error correcting codes that have the ability to fight decoherence is one of the reasons why many believe that large-scale quantum computation and quantum communication will one day be possible  \cite{Shor95,Ste96a,Ste96b,EM96,BDSW96,KL97}.
Quantum information can be protected against noise and decoherence by encoding it into
quantum error-correcting codes, which introduce redundancy to the structure of quantum states.

Quantum stabilizer codes are an extensively analyzed class of quantum error-correcting codes, and  they have many similarities with additive codes from classical error correction theory \cite{CRSS97,CRSS98,Got97,NC00}.
In particular, a quantum code designer can produce quantum stabilizer codes from classical binary and quaternary self-orthogonal codes by means of the CSS and CRSS code constructions, respectively \cite{CS96,Ste96,CRSS97,CRSS98}.

Entanglement-assisted quantum error correction  is a paradigm for
quantum error correction in which a sender and receiver are allowed to share
entanglement before quantum communication begins \cite{BDM06}.
An $[[n,k,d;c]]$ EAQEC code encodes $k$ information qubits into $n$
channel qubits with the help of $c$ pairs of maximally-entangled
Bell states. The code can correct up to $\lfloor \frac{d-1}{2}
\rfloor$ errors acting on the $n$ channel qubits, where  $d$ is
the minimum distance of the code. Standard stabilizer codes are a
special case of EAQEC codes with $c=0$, and we use the notation
$[[n, k, d]]$ for such codes.

It has been shown that EAQEC codes have some advantages over standard stabilizer codes.
For example, only a dual-containing classical linear quaternary code can be transformed
into a standard stabilizer code, but {\it any} classical linear quaternary code can be transformed
into an EAQEC code.
Also,  EA quantum LDPC codes with girth greater than or equal to six can be constructed, and they have good performance \cite{HBD08,HYH11}.
Properties of EAQEC codes and their applications can be found in  Refs.~\cite{WB-2008-77,WB10,WH10,LB10,LB11a}.

The MacWilliams identity for quantum codes connects the weight
enumerator of a classical quaternary self-orthogonal code associated with the quantum code to the weight enumerator of its dual code \cite{SL96,Rains96b,Rains98,AL99}.
This type of MacWilliams identity for quantum stabilizer codes can be directly obtained by applying the Poisson summation formula from the theory of orthogonal groups \cite{LBW13}.
However, the orthogonal group of a stabilizer group with respect to the symplectic inner product (which will be defined below) does not define another quantum stabilizer code.  So this is not a duality between codes in the usual quantum case.

Lai {\it et al}.~defined a duality between entanglement-assisted quantum error-correcting codes based on the theory of orthogonal groups \cite{LBW13}.
This duality is very similar to the classical notion of duality, because the orthogonal group of an EA quantum code also forms a nontrivial EA quantum code.

%In this paper, we provide several contributions that amount to a detailed exposition and extension of our results presented in Ref. \cite{LBW13}:
This paper builds on and extends the results presented in Ref. \cite{LBW13}, giving a more detailed exposition and including additional related topics:
%MAKE THE ORDER CORRESPOND TO ORDER OF RESULTS
\begin{itemize}

\item We demonstrate the duality discussed above with the example of repetition and accumulator EAQEC codes, together with their encoding circuits. Lai and Brun recently constructed a family of EA repetition codes with parameters $[[n,1,n;n-1]]$ for $n$ odd \cite{LB10}.
Herein we produce a family of $[[n,1,n-1;n-1]]$ EA repetition codes for $n$ even, which we prove  to be optimal,\footnote{An $[[n, k, d;c]]$ EAQEC code is optimal in the sense that $d$ is the highest achievable minimum distance for given parameters $n$, $k$, and $c$.}
thus completing the family of EA repetition codes for any $n$.

%\item A quantum analog of the  MacWilliams identity, and the linear programming bound (upper bound) on the minimum distance for EAQEC codes, follow in a natural way by defining duality
%from the theory of orthogonal groups.

\item
By exploiting the linear programming bound derived from the MacWilliams identity for EAQEC codes \cite{LBW13},
 we can now show that several code parameters proposed in Ref.~\cite{LB10} are optimal.

\item We establish the Gilbert-Varshamov bound for EAQEC codes, proving the existence of EAQEC codes with certain parameter values.

%\item We then apply the encoding optimization algorithm from Ref.~\cite{LB10} to find good EAQEC codes with maximal entanglement for $n\leq 15$.
%All of these EAQEC codes have minimum distance greater than or equal to that given by the Gilbert-Varshamov bound.

\item We also derive the quantum version of the Plotkin bound \cite{MS77}, which is tight for codes with small $k$ and maximal entanglement.

\item
The table of upper and lower bounds on the highest achievable
minimum distance of any maximal-entanglement %\footnote{One might wonder why we are considering
%EAQEC codes that exploit the maximum amount of entanglement possible, given that noiseless entanglement could be expensive in practice.
%But there is good reason for doing so. The one-shot father protocol is a random EA quantum code
%\cite{arx2005dev,PhysRevLett.93.230504},
%and it achieves the EA quantum capacity
%of a depolarizing channel (the EA hashing bound \cite{PhysRevLett.83.3081,Bowen02}, see also Section~\ref{sec:hashing bounds} of this paper) by exploiting maximal entanglement. Furthermore,
%there is numerical evidence that maximal-entanglement turbo codes come within a few dB of achieving
%the EA hashing bound \cite{WH10}.}
EAQEC code for $n\leq 15$ in \cite{LBW13}
is established by combining the linear programming bounds \cite{LBW13} with the existence of some EAQEC codes established in this paper.

\item The weight enumerator of a classical code gives an upper bound on the block error probability when transmitting coded bits through a binary symmetric channel \cite{RU08,McE02}.
Since maximal-entanglement EAQEC codes have many similarities with classical codes \cite{LB10,WH10},
we can find an upper bound on the block error probability when transmitting coded quantum information through the depolarizing
channel, and this derivation is similar to the classical derivation \cite{RU08}. We also exploit this result to find an upper bound on the expected block error probability when decoding a random maximal-entanglement EAQEC code.

\item The hashing bound of a quantum channel is an achievable rate for reliable quantum communication \cite{BDSW96}, and it has a simple form for Pauli channels.
We first review a simple proof of the hashing bound for stabilizer codes \cite{Smith06}
and then derive variations of the hashing bound for EAQEC codes.
The proof exploits the method of random stabilizer coding.
\end{itemize}

We organize this paper as follows. We first review the basics of EAQEC codes and give the definition of the dual of an EAQEC code,
and we explain this notion with the example of the dual repetition and accumulator EAQEC codes.
We follow the terminology and notations of EAQEC codes used in Ref.~\cite{LB10}.
For details, we point the reader to Refs.~\cite{BDM06,LB10}.
We review and study the duality,  the MacWilliams identity,  and the linear programming bound for EAQEC codes in Section \ref{sec:duality}.
In Section \ref{sec:bounds}, we  begin with  the Gilbert-Varshamov bound for EAQEC codes.
We  then describe the construction of $[[n,1,n-1;n-1]]$ EA repetition codes for $n$ even and prove other results about the existence of EAQEC codes, including the EA Plotkin bound.
%We finish this section with a table of upper and lower bounds on the minimum distance of any EAQEC code with maximal entanglement for $n\leq 15$.
Section~\ref{sec:error_bound} details an upper bound on the block error probability under maximum-likelihood decoding,
and  Section~\ref{sec:hashing bounds} summarizes variations of hashing bounds for EAQEC codes over Pauli channels.
The final section concludes with a summary and open questions for future research.

\section{Preliminaries}

We begin with some notation.  The Pauli matrices
$$
I=\begin{bmatrix}1 &0\\0&1\end{bmatrix}, \
X=\begin{bmatrix}0 &1\\1&0\end{bmatrix}, \
Y=\begin{bmatrix}0 &-i\\i&0\end{bmatrix}, \
Z=\begin{bmatrix}1 &0\\0&-1\end{bmatrix}
$$
form a basis for the space of linear operators which act on a two-dimensional single-qubit state space.  Let
$$
\mathcal{G}_n=\{e M_1\otimes \cdots \otimes M_n:M_j\in \{I, X, Y, Z\}, e\in \{\pm 1, \pm i \} \}
$$
be the $n$-fold Pauli group.  Any element $g=e M_1\otimes \cdots \otimes M_n\in \mathcal{G}_n$ can be expressed as $g= e' X^{u}Z^{v}$, where  $e'\in \{\pm 1, \pm i \}$; $u=(u_1\cdots u_n)$ and $v=(v_1\cdots v_n)$ are two binary $n$-tuples defined as follows.  If $M_j$ is $I$, $X$, $Z$, or $Y$, then $(u_j, v_j)=(0,0)$, $(1,0)$, $(0,1)$, or $(1,1)$, respectively.  The weight $\wt{g}$ of $g$ is the number of operators  $M_j$ that are not equal to the identity operator $I$.  We use the notation $X_j$, $Y_j$, or $Z_j$ to denote a Pauli operator which acts on qubit number $j$.  Since the overall phase of a quantum state is not important, we consider the quotient of the Pauli group by its center
$\bar{\mathcal{G}}_n= \mathcal{G}_n/\{\pm 1, \pm i\}$, which is an Abelian group and can be generated by a set of $2n$ independent generators.  For $g_1=  X^{u_1}Z^{v_1}$, $g_2= X^{u_2}Z^{v_2} \in \bar{\mathcal{G}}_n$,  the symplectic inner product $*$ in $ \bar{\mathcal{G}}_n $ is defined by
\[
g_1*g_2= u_1\cdot v_2+u_2\cdot v_1 \mod{ 2},
\]
where $\cdot$ is the usual  inner product for binary $n$-tuples.  Note that $*$ is commutative.  We define a map $\phi: \mathcal{G}_n\rightarrow \bar{\mathcal{G}}_n$ by $\phi\left(e X^{u}Z^{v}\right)=  X^{u}Z^{v}$.  For $g,h\in \mathcal{G}_n$, $\phi(g)*\phi(h)=0$ if $g$ and $h$ commute, and  $\phi(g)*\phi(h)=1$, otherwise.  The orthogonal group of a subgroup $V$ of $\bar{\mathcal{G}}_n$ with respect to $*$ is
$$
V^{\perp}= \{ g\in \bar{\mathcal{G}}_n: g*h=0, \forall h\in V \}.
$$
For example, consider a stabilizer subgroup $\mathcal{S}$ of $\mathcal{G}_n$ and its normalizer group $N(\mathcal{S})$.  Then the orthogonal group of $\phi(\mathcal{S})$ is $(\phi(\mathcal{S}))^{\perp}= \phi( N(\mathcal{S}) )$.

An $[[n,k,d]]$ stabilizer code is a $2^k$-dimensional subspace of the $n$-qubit Hilbert space $\mathcal{H}^{\otimes n}$,
and is the joint $+1$-eigenspace of $n-k$ independent generators of a stabilizer subgroup $\mathcal{S}$ of $\bar{\mathcal{G}}_n$.
The minimum distance $d$ is the minimum weight of any element in $\phi(N(\mathcal{S}))\setminus \phi(\mathcal{S})$.

We review some basics of EAQEC codes \cite{BDM06,LB10}.  Suppose Alice and Bob share $c$  maximally-entangled pairs $|\Phi_{+}\rangle_{AB}=\frac{1}{\sqrt{2}}\left(  |00\rangle+|11\rangle\right)$.
 Such a shared pair is called an \emph{ebit}. Alice encodes a $k$-qubit state $|\phi\rangle$  by using an $\left[  \left[  n,k,d;c\right]  \right]$ EAQEC\ code with a \emph{Clifford encoder} $U$, and then sends her $n$ qubits ($k$ information qubits, $n-k-c$ ancilla qubits, and $c$ ebits) to Bob.
 A Clifford operator is a  unitary operator that maps elements $\bar{\mathcal{G}}_n$ to elements of $\bar{\mathcal{G}}_n$ under unitary conjugation.
Alice applies this encoder to her $k$ information qubits, $c$ shares of entangled pairs, and $n-k-c$ ancilla qubits prepared in the state $\ket{0}$, for a total of $n$ qubits on her side.
We assume that Bob's qubits suffer no errors since they do not pass through the noisy channel.
(The minimum distance $d$ will be defined later.)
%An EAQEC code is said to be with maximal entanglement if $n=k-c$.
Suppose the initial state is $\ket{\phi}\ket{\Phi_+}_{AB}^{\otimes c}\ket{0}^{\otimes n-k-c}$.  Let $g_j= UZ_j U^{\dag}$ and $h_j=UX_jU^{\dag}$ for $j=1, \cdots, n$ in $\bar{\mathcal{G}}_n$.  The encoded state $U\ket{\phi}\ket{\Phi_+}_{AB}^{\otimes c}\ket{0}^{\otimes n-k-c}$ has a set of stabilizer generators
\begin{align*}
\{&g_{k+1}^A\otimes Z_{k+1}^B,\cdots,g_{k+c}^A\otimes Z_{k+c}^B, g_{k+c+1}^A\otimes I^B,\cdots,g_{n}^A\otimes I^B,
  h_{k+1}^A\otimes X_{k+1}^B,\cdots,h_{k+c}^A\otimes X_{k+c}^B \}
\end{align*}
in $\bar{\mathcal{G}}_{n+c}$, where the superscript $A$ or $B$ indicates that the operator acts on the qubits of Alice or Bob, respectively.
%In the entanglement-assisted paradigm, we assume that Bob's halves of the ebits do not undergo errors, and we only consider errors acting on Alice's qubits when they pass through the channel.
The case where noise occurs on the ebits was considered in Refs.~\cite{SWOKL08,WH10,LB11a}.

The simplified stabilizer subgroup $\mathcal{S}'$  of $\bar{\mathcal{G}}_n$ is %of the EAQEC code is
%An EAQEC code is defined by the simplified stabilizer group $\mathcal{S}'$ of the encoded state:
\[
\mathcal{S}' = \langle g_{k+1}, \cdots, g_{k+c}, h_{k+1},
\cdots, h_{k+c}, g_{k+c+1}, \cdots, g_{n} \rangle.
\]
%$\mathcal{S}'=$ $\langle g_{k+1}$, $\cdots,$ $g_{k+c}$,
%$h_{k+1}$, $\cdots,$ $h_{k+c}$, $g_{k+c+1}$, $\cdots,$ $g_{n} \rangle$.
Note that the commutation relations are as follows:
\begin{align}
&g_i*g_j=0\mbox{ for $i\neq j$}, \label{eq:commutation_1}\\
&h_i*h_j=0\mbox{ for $i\neq j$}, \label{eq:commutation_2}\\
&g_i*h_j=0 \mbox{ for $i\neq j$}, \label{eq:commutation_3}\\
&g_i*h_i=1 \mbox{ for all $i$}. \label{eq:commutation_4}
\end{align}
We say that $g_i$ and $h_i$ are  \emph{symplectic partners} for $i=1, \cdots, k+c$.
%meaning that they anticommute with each other and they commute with
The logical subgroup $\mathcal{L}$ of $\bar{\mathcal{G}}_n$ of the encoded state is
$$\mathcal{L}= \langle g_1, \cdots, g_k, h_1, \cdots, h_k \rangle.$$
The symplectic subgroup $\mathcal{S}_S$ of $\mathcal{S}'$  is the subgroup generated by the $c$ pairs of symplectic partners of $\mathcal{S}'$:
$$\mathcal{S}_S= \langle g_{k+1}, \cdots, g_{k+c}, h_{k+1}, \cdots, h_{k+c}\rangle$$
The isotropic subgroup $\mathcal{S}_I$ of $\mathcal{S}'$ is the subgroup generated by the generators $g_i$ of $\mathcal{S}'$ such that
$g_i*g=0$ for all $g$ in $\mathcal{S}'$:
$$\mathcal{S}_I =\langle g_{k+c+1}, \cdots, g_{n}\rangle.$$
Notice that  $\mathcal{S}'= \mathcal{S}_S\times \mathcal{S}_I$  in  $\bar{\mathcal{G}}_n$.
%to mean that $\mathcal{S}'$ is generated by the generators of $\mathcal{S}_S$ and $\mathcal{S}_I$.
The minimum distance $d$ of the EAQEC code is the minimum weight of any element in
$(\phi(\mathcal{S}'))^{\perp}\setminus \phi(\mathcal{S}_I)$.

%where $\mathcal{L}\times \mathcal{S}_I$ and
% $\mathcal{S}_S\times \mathcal{S}_I$  are dual to each other.

\section{Duality in Entanglement-Assisted Quantum Codes}
\label{sec:duality}
\subsection{The Dual of an Entanglement Assisted Quantum Code}
%
%Observe that the orthogonal group of $\mathcal{S}'= \mathcal{S}_S\times \mathcal{S}_I$ in $\bar{\mathcal{G}}_n$ is $\mathcal{L}\times \mathcal{S}_I$, which defines another EAQEC code with symplectic subgroup $\mathcal{L}$ and isotropic subgroup $\mathcal{S}_I$.  The number of independent generators of $ \mathcal{S}'= \mathcal{S}_S\times \mathcal{S}_I $ is  $K= 2c+(n-k-c)=n-k+c$, and the number of independent generators of its orthogonal group  $\mathcal{L}\times \mathcal{S}_I$ is  $K'=2k+ (n-k-c)=n+k-c $.  These parameters satisfy the following relation:
%\[
%K+K'=2n=N,
%\]
%where $N$ is the number of independent generators for the full quotient group $\bar{\mathcal{G}}_n$.  This equation corresponds to the classical duality between a code and its dual code, which motivates the definition of the dual code of an EAQEC code as follows.

%\bd
The dual of an $[[n,k,d;c]]$ EAQEC code, defined by a simplified stabilizer group $\mathcal{S}'=\mathcal{S}_S\times \mathcal{S}_I$ and a logical group $\mathcal{L}$, is the $[[n,c,d';k]]$ EAQEC code with  $\mathcal{L}\times \mathcal{S}_I$ being the simplified stabilizer group and $\mathcal{S}_S$  being the logical group for some minimum distance $d'$ \cite{LBW13}.
When $c=0$, the code is a standard stabilizer code. This case is not a concern of this paper.

%, with a stabilizer group $\mathcal{S}= \mathcal{S}_I=\langle g_{k+1}, \cdots, g_{n}\rangle$, and a logical group $\mathcal{L}=\langle g_1, \cdots, g_k, h_1, \cdots, h_k \rangle$.  $\mathcal{S}_S$ is the trivial group in this case.
%The dual code of the standard stabilizer code is an $[[n,0,d';k]]$ EAQEC code defined by the simplified stabilizer group  $\mathcal{L}$---that is, a single stabilizer state that does not encode any quantum information.
%This suggests that the standard paradigm
When $c=n-k$, we call such a code a {\it maximal-entanglement} EAQEC code.  In this case, $\mathcal{S}_I$ is the trivial group that contains only the identity, and the simplified stabilizer group is $\mathcal{S}_S$.  Its dual code is a maximal-entanglement EAQEC code  defined by the logical group $\mathcal{L}$.
\begin{figure*}
[ptb]
\begin{center}
\includegraphics[
natheight=4.760800in, natwidth=12.773300in,
width=5.305518in
]%
{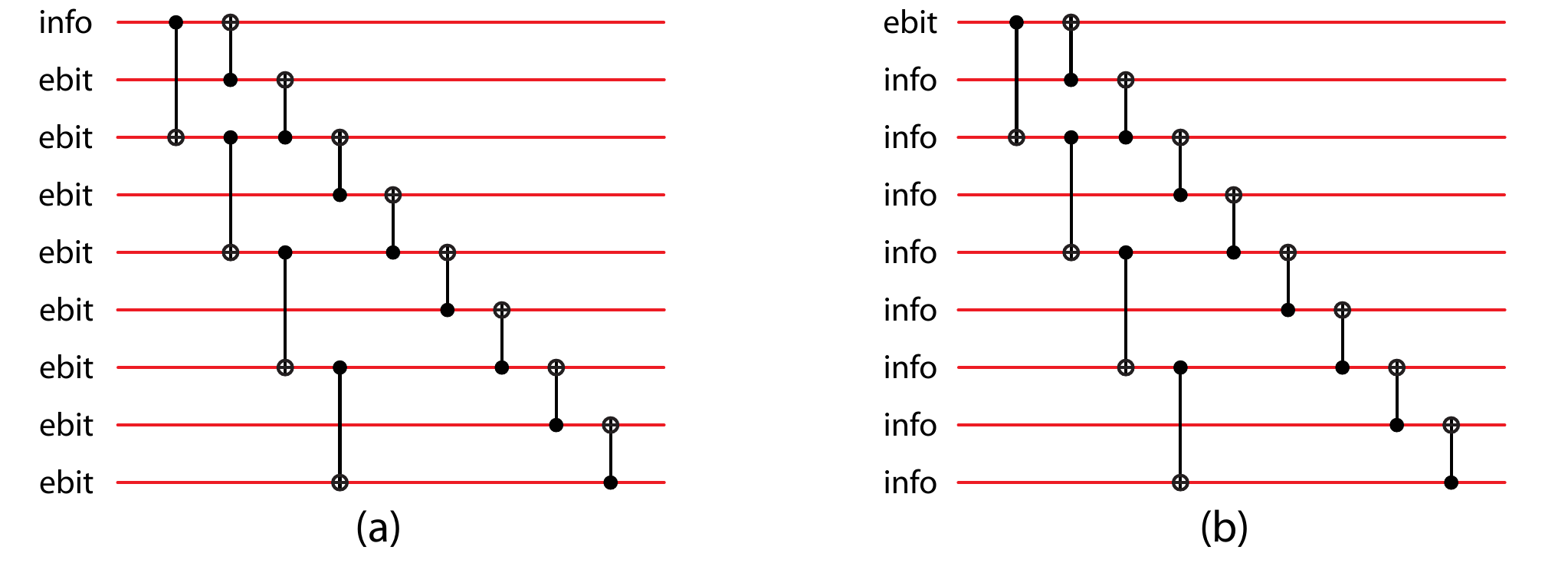}%
\caption{(a) The encoder for a $\left[  \left[  9,1,9;8\right]
\right]  $ EA repetition code consists of a
periodic cascade of CNOT gates. The encoder for arbitrary $\left[
\left[  n,1,n;n-1\right]  \right]  $ EA
repetition codes extends naturally from this design. We can
implement these encoders with a simple quantum shift-register circuit
that uses only one memory qubit \cite{W09}. (b)
Considering the circuit in (a) but changing the information qubit
to an ebit and all of the ebits to information qubits gives the
encoder for the dual of the EA repetition code,
namely, the $\left[  \left[ 9,8,2;1\right]  \right]  $
EA accumulator code. This circuit naturally
extends to encode the $\left[  \left[  n,n-1,2;1\right]
\right]  $ EA accumulator codes. Simple variations of
the above circuit encode the even $n$ repetition and accumulator EAQEC codes,
and we discuss them in Section~\ref{sec:max-ent-rep-acc}.}%
\label{fig:rep-acc}%
\end{center}
\end{figure*}
%EndExpansion
As an example, consider the  class of $\left[
\left[  n,1,n;n-1\right]  \right]  $\ EA
repetition codes for $n$ odd \cite{LB10}.
The two logical operators for the one logical qubit of this code are as follows:%
\begin{equation}%
\begin{array}
[c]{cccccc}%
X & X & X & \cdots & X & X\\
Z & Z & Z & \cdots & Z & Z
\end{array}
\label{eq:log-ops-EA-rep}
\end{equation}
%These logical operators are directly imported from the generator matrix in (\ref{eq:repetition-generator}).
The simplified stabilizer generators are as follows:%
\begin{equation}%
\begin{array}
[c]{cccccc}%
Z & Z & I & \cdots & I & I\\
I & Z & Z & \cdots & I & I\\
\vdots &  & \ddots & \ddots &  & \vdots\\
I & I & I & \cdots & Z & Z\\
%\end{array}
%\end{equation*}%
%\begin{equation}%
%\begin{array}
%[c]{cccccc}%
X & X & I & \cdots & I & I\\
I & X & X & \cdots & I & I\\
\vdots &  & \ddots & \ddots &  & \vdots\\
I & I & I & \cdots & X & X
\end{array}
\label{eq:EA-stab-ops}
\end{equation}
One can determine %which operators should go on Bob's side
the symplectic pairs
by
performing a symplectic Gram-Schmidt orthogonalization of the
above operators \cite{wilde-2008}.
%The claim is that the dual of this code is an $\left[  \left[ n,n-1,2;1\right]  \right]  $ accumulator code. We obtain this code
If we interchange the roles of the stabilizer subgroup and the logical operator subgroup,
we obtain an $\left[  \left[ n,n-1,2;1\right]  \right]  $ EA accumulator code.

To make this more precise, consider the encoding circuits in
Figure~\ref{fig:rep-acc}. The circuit in
Figure~\ref{fig:rep-acc}(a) is the encoder of a $\left[  \left[
9,1,9;8\right]  \right]  $ EA repetition code.
Swapping the information qubit for an ebit and all of the ebits
for information qubits gives the encoder of
Figure~\ref{fig:rep-acc}(b), which encodes a $\left[  \left[
9,8,2;1\right]  \right]  $ EA accumulator code.
To illustrate that the circuit is working as expected, let us
consider it acting on the first five qubits only (just to
simplify the analysis). Inputting the following two
operators at the information qubit slot%
\[%
\begin{array}
[c]{ccccc}%
X & I & I & I & I\\
Z & I & I & I & I
\end{array}
\]
gives the following two logical operators:%
\[%
\begin{array}
[c]{ccccc}%
X & X & X & X & X\\
Z & Z & Z & Z & Z
\end{array}
,
\]
and these operators match the form of the logical operators for
the repetition code in (\ref{eq:log-ops-EA-rep}). Inputting the
following operators at the
ebit slots%
\[%
\begin{array}
[c]{ccccc}%
I & X & I & I & I\\
I & Z & I & I & I\\
I & I & X & I & I\\
I & I & Z & I & I\\
I & I & I & X & I\\
I & I & I & Z & I\\
I & I & I & I & X\\
I & I & I & I & Z
\end{array}
,
\]
gives the following operators%
\[%
\begin{array}
[c]{ccccc}%
X & X & I & I & I\\
I & Z & Z & Z & Z\\
I & X & X & X & X\\
Z & Z & I & I & I\\
I & I & X & X & I\\
I & I & I & Z & Z\\
I & I & I & X & X\\
I & I & Z & Z & I
\end{array}
,
\]
which we can transform by row operations to be the same as the
operators in (\ref{eq:EA-stab-ops}).

\subsection{The MacWilliams Identity} \label{sec:MacWilliams}

The MacWilliams identity for general quantum codes can be obtained by applying the Poisson summation formula  from the theory of orthogonal groups \cite{Knapp06,LBW13}.
We restate Theorem 2 and Corollary 3 in \cite{LBW13} here.

\begin{theorem}[Theorem 2 of \cite{LBW13}] \label{thm:MacWilliamsIdentityOrthogonalGroup}
Suppose $$W_V(x,y)= \sum_{w=0}^n B_w x^{n-w} y^w$$ and
$$W_{V^{\perp}}(x,y)= \sum_{w'=0}^n A_{w'} x^{n-w'} y^{w'}$$ are the weight enumerators of    subgroup $V$ of $\bar{\mathcal{G}}_n$ and  its orthogonal group $V^{\perp}$  in $\bar{\mathcal{G}}_n$, respectively.  Then
\begin{align}
W_{V}(x,y) =\frac{1}{|V^{\perp}|} W_{V^{\perp}}(x+3y,x-y),
\label{eq:MacWilliamsIdentityOrthogonalGroup}
\end{align}
or
\begin{align}
B_w=\frac{1}{|V^{\perp}|} \sum_{w'=0}^n
P_w(w',n) A_{w'},
\ \mbox{ for $w=0, \cdots, n$},
 \label{eq:MacWilliamsIdentityCoeff}
\end{align}
where $
P_w(w',n)= \sum_{u=0}^{w}(-1)^u 3^{w-u} {w'\choose u}{n-w'\choose
w-u}
$
is the Krawtchouk polynomial \cite{MS77}.
\end{theorem}
\begin{corollary}[Corollary 3 of \cite{LBW13}] \label{thm:GeneralQuantumMacWilliamsEAQEC}
%Suppose the stabilizer group of an EAQEC code is $\mathcal{S}'$ and its logical group is $\mathcal{L}$.
%Suppose the weight enumerators of $\mathcal{S}'={\mathcal{S}_S \times
%\mathcal{S}_I}$  and ${\mathcal{L} \times \mathcal{S}_I}$ are
%$W_{\mathcal{S}_S\times \mathcal{S}_I}(x,y)=\sum_{w=0}^n A_w x^{n-w}y^w$ and $W_{\mathcal{L}\times \mathcal{S}_I}(x,y)=\sum_{w=0}^n B_w x^{n-w}y^w$, respectively.
The MacWilliams identities for EAQEC codes are as follows:
\begin{align}
W_{\mathcal{L}\times \mathcal{S}_I}(x,y)& =\frac{1}{|\mathcal{S}_S\times \mathcal{S}_I|}  W_{\mathcal{S}_S\times \mathcal{S}_I}(x+3y,x-y), \label{eq:MI EAQEC1} \\
W_{ \mathcal{S}_I}(x,y)& =\frac{1}{|\mathcal{L}\times \mathcal{S}_S\times \mathcal{S}_I|}  W_{\mathcal{L}\times \mathcal{S}_S\times \mathcal{S}_I}(x+3y,x-y).\label{eq:MI EAQEC2}
\end{align}
%or
%\[
%B_w=\frac{1}{|\mathcal{S}_S\times \mathcal{S}_I|} \sum_{w'=0}^n
%P_w(w',n) A_{w'},
%\ \mbox{ for $w=0, \cdots, n$},
%\]

\end{corollary}

In the case of maximal-entanglement EAQEC codes, $\mathcal{S}_I$ is the trivial group and there is no degeneracy.
%Assume that the weight enumerators of $\mathcal{S}'$ and $\mathcal{L}$ are respectively  $C(z)= \sum_{w=0}^n C_w z^w$ and $D(z)= \sum_{w=0}^n D_w z^w$,
%where \[C_w=\sum_{E\in \mathcal{S}'_w}1= |\mathcal{S}'_w|,\
%\ D_w=\sum_{E\in \mathcal{L}_w}1= |\mathcal{L}_w|. \]
%and $E_w$ is an element in $\mathcal{G}_n$ of weight $w$.
%the number of elements of $\mathcal{S}'$ and $\mathcal{L}$, respectively.
%\bl \label{lemma:half_commute}
%For an operator $E\in \mathcal{G}^n$, the Pauli group of $n$-qubit,
%\begin{align}
%\sum_{M\in \mathcal{S}'} (-1)^{[ E, M ]}= \left\{%
%\begin{array}{ll}
%    |\mathcal{S}'|, & \hbox{ if $E\in \mathcal{L}$;} \\
%    0, & \hbox{if $E\notin \mathcal{L}$,} \\
%\end{array}%
%\right.  \label{eq:half_commute}
%\end{align}
%where
%\[
%[ E,M ] = \left\{%
%\begin{array}{ll}
%0, & \hbox{ if $E$ and $M$ commute;} \\
%    1, & \hbox{ if $E$ and $M$ anti-commute.} \\
%\end{array}%
%\right.
%\]
%\el
%\bt \label{thm:QuantumMacWilliamsEAQEC}
%If the weight enumerators of $\mathcal{S}'$  and $\mathcal{L}$ are $C(z)$ and $D(z)$, respectively, then the MacWilliams identities of an EAQEC code with maximal ebits are
%\begin{align}
%D_w=& \frac{1}{|\mathcal{S}'|} \sum_{w'=0}^n \sum_{u=0}^w(-1)^u 3^{w-u}
%{w'\choose u}{n-w'\choose w-u}C_{w'}, \label{eq:QuantumMacWilliamsEAQEC}
%\end{align}
%for $w=0,\cdots, n$, or
%\begin{align*}
%B(z)=\frac{1}{|\mathcal{S}'|}(1+3z)^n A(\frac{1-z}{1+3z}).
%\end{align*}
%\et
If we exchange the roles of $\mathcal{S}_S$ and $\mathcal{L}$, we obtain an $[[n,n-k,d'; k]]$ EAQEC code, which is the dual of the original  $[[n,k,d; n-k]]$ EAQEC code.
The minimum distance $d'$ of this $[[n,n-k,d'; k]]$ EAQEC code is the minimum weight of a nontrivial element in $\mathcal{L}^{\perp}=\mathcal{S}_S$.
Thus $d'$  can be determined from the MacWilliams identity and the weight enumerator of the $[[n,k,d; n-k]]$  EAQEC code, as in the following example.
%We have \[C_0=D_0=1.\]
%\[C_w\geq 0, D_w \geq 0.\]
\be
The dual of the $[[n,1,n;n-1]]$ repetition code  is the $[[n,n-1,2;1]]$ accumulator code whenever $n$ is odd.
The coefficients of $W_{\mathcal{L}}(x,y)=\sum_{w=0}^n B_w x^{n-w}y^{w}$ for the odd-$n$ $[[n,1,n;n-1]]$ repetition code are $B_{(n)}\triangleq (B_0,\cdots,B_n)=(1, 0,0,\cdots, 0, 3)$.
Let $A_{(n)}= (A_0,\cdots, A_n)$.
Using the MacWilliams identity, we obtain the weight enumerators $W_{\mathcal{S}_S}(x,y)=\sum_{w=0}^n A_w x^{n-w}y^{w}$  of these dual EAQEC codes:
\begin{align*}
A_{(3)}=&(1,     0,     9,     6),\\
B_{(3)}=&(1,     0,     0,     3),\\
A_{(5)}=&(1,   0,   30,  60,  105, 60), \\
B_{(5)}=&(1,0,0,0,0,3),\\
A_{(7)}=&(1,   0,   63,  210, 735, 1260,    1281,    546),\\
B_{(7)}=&(1,0,0,0,0,0,0,3), \\
A_{(9)}=&(1,   0,   108, 504, 2646,    7560,    15372,  \\
& 19656,   14769,   4920),\\
 B_{(9)}=&(1,0,0,0,0,0,0,0,0,3),\\
A_{(11)}=&(1, 0, 165, 990, 6930, 27720, 84546, \\
&  180180, 270765, 270600, 162393, 44286),\\
B_{(11)}=&(1,0,0,0,0,0,0,0,0,0,0,3) .
\end{align*}
%We have also verified the correctness of the above weight enumerators by counting the logical operators with a computer program.
\ee

\subsection{Linear Programming Bounds for EAQEC Codes}\label{sec:linear programming}
The significance of the MacWilliams identities  is that linear
programming techniques  can be applied to find upper bounds on the
minimum distance of EAQEC codes \cite{LBW13}. %We first provide the linear
We have the MacWilliams identities (\ref{eq:MI EAQEC1}) and (\ref{eq:MI EAQEC2}) in Corollary \ref{thm:GeneralQuantumMacWilliamsEAQEC}.
Suppose the weight
enumerators of   $\mathcal{S}_S\times \mathcal{S}_I$,
$\mathcal{L}\times \mathcal{S}_I$,  $\mathcal{S}_I$, and
$\mathcal{L}\times \mathcal{S}_S\times  \mathcal{S}_I$ are $W_{\mathcal{S}_S\times \mathcal{S}_I}(x,y)=\sum_{w=0}^n A_w x^{n-w}y^{w}$, $W_{\mathcal{L}\times \mathcal{S}_I}(x,y)=\sum_{w=0}^n B_w x^{n-w}y^{w}$,
$W_{ \mathcal{S}_I}(x,y)=\sum_{w=0}^n C_w x^{n-w}y^{w}$, and $W_{\mathcal{L}\times \mathcal{S}_S\times \mathcal{S}_I} (x,y)$ $= \sum_{w=0}^n D_w x^{n-w}y^{w}$,
respectively.
%Suppose  %the weight enumerator of $\mathcal{S}_I$ be $C(z)$ and
%the weight enumerators of are  respectively.

Since the minimum distance of an EAQEC code is the minimum weight of any element in $(\mathcal{L}\times
\mathcal{S}_I) \backslash \mathcal{S}_I$, it is the minimum nonzero integer $w$ such that $B_w-C_w>0$.
With constraints on $B_w$'s and  $C_w$'s, we can find the linear
programming bound on the minimum distance of the EAQEC code. To sum up, we have the
following constraints: {
%\small
%\footnotesize
%\scriptsize
\begin{align*}
&A_0=B_0=C_0=D_0=1;\\
&A_w\geq 0, B_w \geq 0, C_w\geq 0, D_w \geq 0, \mbox{ for $w=1, \ldots, n$};\\
&A_w\leq |\mathcal{S}_S\times \mathcal{S}_I|, B_w \leq |\mathcal{L}\times \mathcal{S}_I|, \mbox{ for $w=1, \ldots, n$};\\
&C_w\leq |\mathcal{S}_I|, D_w \leq | \mathcal{L}\times \mathcal{S_S}\times\mathcal{S}_I|,\mbox{ for $w=1, \ldots, n$};\\
&D_w\geq A_w, D_w\geq B_w, D_w\geq C_w,  \mbox{ for $w=1, \ldots, n$};\\
&A_w \geq C_w, B_w \geq C_w, \mbox{ for $w=1, \ldots, n$};\\
%\end{align*}
%\begin{align*}
&\sum_{w=0}^n A_w= |\mathcal{S}_S\times \mathcal{S}_I|,\ \sum_{w=0}^n B_w =|\mathcal{L}\times \mathcal{S}_I|,\\
&\sum_{w=0}^n C_w= |\mathcal{S}_I|, \ \sum_{w=0}^n D_w =|\mathcal{L}\times  \mathcal{S_S}\times \mathcal{S}_I|;\\
&B_w=\frac{1}{|\mathcal{S}_S\times \mathcal{S}_I|} \sum_{w'=0}^n
P_w(w',n) A_{w'},
\ \mbox{ for $w=0, \ldots, n$};\\
&D_w=\frac{1}{| \mathcal{S}_I|} \sum_{w'=0}^n  P_w(w',n) C_{w'},
\ \mbox{ for $w=0, \ldots, n$};\\
%\end{align*}
%\begin{align*}
&B_w=C_w,   \mbox{ for $w=1, \ldots, d-1$};
%&B_w\geq C_w,   \mbox{ for $w=d, d+1 \cdots, n$};
\end{align*}
}

%There is no linear function to be optimized in this coding problem.
If we cannot find any solutions to the integer
programming problem with the above constraints for a certain $d$,
this result implies that there is no $[[n,k,d; c]]$ EAQEC code. If
$d^*$ is the smallest such $d$, then $d^*-1$ is an upper
bound on the minimum distance of an $[[n,k,d;c]]$ EAQEC code. This
is the linear programming bound for EAQEC codes with $0<c<n-k$. If
we replace the constraints
\begin{align*}
&B_w=C_w,   \mbox{ for $w=1, \ldots, d-1$},
%&B_w\geq C_w,  \mbox{ for $w=d, d+1 \cdots, n$};
\end{align*}
with
\begin{align*}
&A_w=C_w,   \mbox{ for $w=1, \ldots, d-1$},
%&A_w\geq C_w,   \mbox{ for $w=d, d+1 \cdots, n$};
\end{align*}
this gives the constraints of the linear programming bound on the
minimum distance of the $[[n,c,d;k]]$ dual code.

\be Consider the  $[[8,3,5;5]]$ EAQEC code from the
random optimization algorithm in Ref.~\cite{LB10}. The linear
programming bound shows that there is no $[[8,3,d;5]]$ EAQEC code
with $d>5$, and thus the $[[8,3,5;5]]$ code is optimal.
%\footnote{We used the optimization software LINGO from LINDO Systems to solve the integer programming problem for these examples.}
\ee
\be
Consider the   $[[15,7,6;8]]$ EAQEC code from the random
optimization algorithm in Ref.~\cite{LB10}. The linear programming
bound shows that no $[[15,7,d;8]]$ EAQEC code  with $d>7$ exists;
however, it does not rule out the existence of a $[[15,7,7;8]]$
code.
%Hence an $[[15,7,7;8]]$ code could be found by optimizing over more sets of row operators or over a better code.

\ee
\be
 The linear programming bound of the $[[7,1;4]]$ code is $d\leq 6$, which is the same as the singleton bound.
Hence the $[[7,1,5;4]]$ EAQEC code from the encoding
optimization algorithm in Example 4 of \cite{LB10} does not
achieve the linear programming bound. It follows that the
$[[7,1,5;5]]$ EAQEC code does not achieve the upper bound either. \ee

\be The linear programming bound of the $[[8,3;3]]$ EAQEC code is $d\leq
4$, which improves the singleton bound: $d\leq 5$. Hence the
$[[8,3,4;3]]$ EAQEC code from the encoding optimization
algorithm  in Example 8 of \cite{LB10} is optimal. On the other
hand, the linear programming bound of the $[[8,3;4]]$ EAQEC code is
$d\leq 5$, which is the same as the singleton bound. Hence the
$[[8,3,4;4]]$ EAQEC code does not achieve the upper bound.
\ee

\be The linear programming bounds of $[[9,1;c]]$ EAQEC codes for
$c=3,4,5$  are $d\leq 8$; however, the singleton bounds are $d\leq
6$, $d\leq 7$, and $d\leq 7$, respectively.
%The linear programming bound of a $[[9,1;3]]$ code is $d\leq 8$; however, the singleton bound is $d\leq 6$.
\ee

\begin{remark} From these three examples, we can determine that the linear programming bound
might be better or worse than the Singleton bound when $0 <
c < n - k$.\end{remark}

\section{Bounds on EAQEC Codes} \label{sec:bounds}

Table I of Ref.~\cite{LBW13} establishes
lower and upper bounds on the minimum distance of
maximal-entanglement EAQEC codes with length $n\leq 15$.
This section provides a detailed justification for these results.
We begin by discussing the existence of arbitrary EAQEC
codes, followed by some specific EAQEC code constructions.

%This table is improved in .
%for there is no degeneracy in the case of maximal entanglement.
%Codes that achieve the upper bounds are detailed in Ref.~\cite{LBW13}.

%This section establishes a table of upper and lower bounds on the best minimum distances of maximal-entanglement EAQEC codes for $n\leq 15$.
%boundgive the existence of

The existence of an $[[n,k,d]]$ stabilizer code implies the existence of an $[[n,k,d'\geq d;c]]$ EAQEC code,
since we can replace ancilla qubits with ebits and then optimize the encoding operator \cite{LB10}.
Therefore, the lower bound on the minimum distance of regular stabilizer codes \cite{CRSS98} can be applied here.
Similarly, the existence of an $[[n,k,d;c]]$ EAQEC code where $c<n-k$ implies the existence of an $[[n,k,d'\geq d;c'>c ]]$ EAQEC code.
This establishes the existence of many $[[n,k;c]]$ EAQEC codes.

\subsection{Gilbert-Varshamov Bound for EAQEC Codes}

Consider the stabilizer group $\mathcal{S}$ of an $[[n,k,d;c]]$ EAQEC code, which is a subgroup of the Pauli group~$\mathcal{G}^{n+c}$.
We consider only  error operators in the group $\mathcal{G}^n$ because the entanglement-assisted paradigm assumes  that the ebits on Bob's side of the channel are not subject to errors.
The EAQEC code is defined in an $(n+c)$-qubit space, but only the first $n$ qubits suffer from errors.
Following the argument of the quantum Gilbert-Varshamov bound \cite{EM96}, we obtain
the Gilbert-Varshamov bound for EAQEC codes.
%We write this result as a conjecture, since a strict proof is needed.
However, we will show that there are maximal-entanglement EAQEC codes with minimum distance
higher than this bound for $n\leq 15$.
\begin{theorem}
Given $n,d,c$, let
\[
k= \left\lceil \log_2 \left( \frac{ 2^{n+c} }{\sum_{j=0}^{d-1}3^j {n\choose j}}  \right)\right \rceil.
\]
The Gilbert-Varshamov bound for EAQEC codes states that if $0 \leq
k\leq n-c$, then
 there exists an $[[n,k,d;c]]$ EAQEC code. Equivalently,
\[
\sum_{j=0}^{d-1}3^j {n\choose j} 2^k \geq 2^{n+c}.
\]
\end{theorem}

Most of the lower bounds in Table 1 of Ref.~\cite{LBW13} are slightly higher than the Gilbert-Varshamov bounds for $n\leq 15$.

\subsection{Maximal-Entanglement EAQEC Repetition and Accumulator Codes for Even $n$}
\label{sec:max-ent-rep-acc}

%\subsection{ Entanglement-Assisted Repetition Codes}
Lai and Brun proposed a construction of $[[n, 1,n;n-1]]$
EA repetition codes for $n$ odd in Ref.~\cite{LB10}.
In the case of even $n$, that construction gives a series of $[[n,0,n;n-2]]$ EA repetition codes with no information qubits.
%\bt \label{thm:new code}
%    There are  entanglement-assisted repetition codes for  $n$ even.
%\et
%(The two logical operators for the one logical qubit of this code are as follows:%
%\begin{align}
%& X\otimes X\otimes X\otimes\cdots\otimes X\otimes X,\nonumber\\
%& Z\otimes Z\otimes Z\otimes\cdots\otimes Z\otimes Z.\label{eq:log-ops-EA-rep}%
%\end{align}
%These logical operators are the direct import of the generator matrix in
%(\ref{eq:repetition-generator}). The \textquotedblleft
%simplified\textquotedblright\ check matrix is as follows:%
%\begin{gather}
%Z\otimes Z\otimes I\otimes\cdots\otimes I\otimes I,\nonumber\\
%I\otimes Z\otimes Z\otimes\cdots\otimes I\otimes I,\nonumber\\
%\vdots\nonumber\\
%I\otimes I\otimes I\otimes\cdots\otimes Z\otimes Z,\nonumber\\
%X\otimes X\otimes I\otimes\cdots\otimes I\otimes I,\nonumber\\
%I\otimes X\otimes X\otimes\cdots\otimes I\otimes I,\nonumber\\
%\vdots\nonumber\\
%I\otimes I\otimes I\otimes\cdots\otimes X\otimes X.\label{eq:EA-stab-ops}%
%\end{gather}
%One can determine which operators should go on Bob's side by performing a
%symplectic Gram-Schmidt orthogonalization of the above operators.)

In this section, we construct $[[n, 1,n-1;n-1]]$
EA repetition codes for $n$ even. We obtained these codes both with
the techniques from Ref.~\cite{LB10} and by realizing that the encoding circuit in Figure~\ref{fig:rep-acc} can encode them.
The duals of these codes are the $[[n, n-1, 1; 1]]$ EA accumulator codes for even $n$.
Theorems~\ref{thm:ne-rep} and \ref{thm:ne-acc} show that both of these code constructions are optimal, in the sense that
 $[[n,1,n;n-1]]$ and $[[n,n-1,2;1]]$ EAQEC codes do not exist for even $n$.
 Thus, the results here complete our understanding of the dual classes of EA repetition and accumulator codes for arbitrary $n$.
\begin{theorem} \label{thm:even repetition}
    There are $[[n,1,n-1;n-1]]$ EA repetition codes for  $n$ even.
    The duals of these codes are the $[[n,n-1,1;1]]$ EA accumulator codes.
\end{theorem}
\begin{proof} Suppose $H_{(n-1)}$ is an $(n-2)\times (n-1)$ parity-check matrix of a classical $[n-1,1,n-1]$ binary repetition code:  % and % and $O_{i\times j}$ is an $i\times j$ zero matrix.
\[
H_{(n-1)}= \begin{bmatrix}
1&1&0&\cdots &0 &0\\
0&1&1&\cdots &0 &0\\
\vdots& &\ddots&\ddots&&\vdots\\
%0&\cdots &\cdots & 1&1&0\\
0&\cdots &\cdots & 0&1&1
\end{bmatrix}.
\]
We define two  $(n-1)\times n$  matrices
\[
H_1=
\begin{bmatrix}
%0&&&&\\
0&&&&\\
\vdots &&H_{(n-1)}&&\\
0&&&&\\
1&1&\cdots&1&0\\
\end{bmatrix},
\]
and
\[
H_2=
\begin{bmatrix}
%&&&&0\\
&&&&0\\
&&H_{(n-1)}&&\vdots\\
&&&&0\\
1&1&\cdots&1&1\\
\end{bmatrix}.
\]
Consider a simplified check matrix of the form
%\begin{align*}
$$H'=\begin{bmatrix}
O & H_1\\
H_2 & O\\
\end{bmatrix}.
$$ %\end{align*}
Consider the matrix $H_1H_2^T$. We have that
\[
\left[H_1 H_2^T\right]_{i,j}=\left\{%
\begin{array}{ll}
    1, &  \begin{array}{c}
\text{if } i=j \mbox{ for } j=1,\cdots, n-1, \mbox{ or }\\
 i=j-2 \mbox{ for } j=3,\cdots, n-2\\
\end{array}
\\
    0, & \hbox{else.} \\
\end{array}%
\right.
\]
For example, when $n=6$,
\[
H_1H_2^T= \begin{bmatrix}
1&0&1&0& 0\\
0&1&0&1& 0\\
0&0&1&0& 0\\
0&0&0&1& 0\\
0&0&0&0&1
\end{bmatrix}.
\]
Thus the number of symplectic pairs in $H'$ is as follows \cite{WB-2008-77}:
\[
\frac{1}{2}\mbox{rank}(H'\Lambda H'^T)=\mbox{rank}(H_1 H_2^T)=n-1.
\]
%since
%\[
%H_1H_2^T= \begin{bmatrix}
%1&0&1&0&\cdots &0&0& 0\\
%0&1&0&1&\cdots &0&0& 0\\
%0&0&1&0&\ddots &0&0& 0\\
%\vdots&\vdots&&\ddots &\ddots&\ddots&\vdots&\vdots\\
%0&0 &&\cdots& &0& 1&0\\
%0&0 &&\cdots& &1& 0&0\\
%0&0 &&\cdots& &0& 1&0\\
%0&0 &&\cdots& &0& 0&1
%\end{bmatrix}.
%\]
The simplified logical matrix is
\[
L'=\left[
\begin{array}{c|c}
00\cdots  00&11\cdots  1\\
11\cdots  10&00\cdots  0\\
\end{array}
\right],
\]
which implies the minimum distance is $n-1$.
Therefore, $H'$ and $L'$ define an $[[n,1,n-1;n-1]]$ EAQEC code.
One obtains the dual $[[n,n-1,1;1]]$ codes simply by swapping the roles of the logical matrix
and the simplified check matrix.

\end{proof}
\noindent This completes the family of EA repetition and accumulator codes for any $n$.
The encoding circuit of Figure \ref{fig:rep-acc} encodes these even-$n$ repetition codes
with the exception that the last qubit is removed, the last CNOT in the first string does not act, and the last CNOT in the second string does not act.

\be
The coefficients of $W_{\mathcal{L}}(x,y)=\sum_{w=0}^n B_w x^{n-w}y^{w}$ for the even-$n$ $[[n,1,n-1;n-1]]$  EA repetition code are $B_{(n)}=(1,0,\cdots, 0, 1, 2)$.
Using the MacWilliams identity, we obtain the weight enumerators $W_{\mathcal{S}_S}(x,y)=\sum_{w=0}^n A_w x^{n-w}y^{w}$ of these dual even-$n$ EAQEC codes:
\begin{align*}
A_{(4)}=&(1,     1,    15 ,   27 ,   20), \\
B_{(4)}=&(1,     0, 0, 1,     2),\\
A_{(6)}=&(1,     1,    40 ,  130 ,  305 ,  365 ,  182), \\
B_{(6)}=&(1,0,0,0,0,1,2),\\
A_{(8)}=&(1 , 1, 77, 357, 1435, 3395, 5103, 4375, 1640),\\
B_{(8)}=&(1,0,0,0,0,0,0,1,2), \\
A_{(10)}=&(1, 1, 126, 756, 4326, 15246, 38304, 65604,\\
    &  73809, 49209, 14762),\\
 B_{(10)}=&(1,0,0,0,0,0,0,0,0,1,2),\\
A_{(12)}=&(1, 1, 187, 1375, 10230, 47850, 168630, 432894,\\
& 811965, 1082565, 974303, 531443, 132860),\\
 B_{(12)}=&(1,0,0,0,0,0,0,0, 0, 0, 0, 1,2).
\end{align*}
\ee
% that you do not execute the last CNOT gate in the first string, and the last CNOT gate in the second string (so that the gate acts on an even number of qubits).

The even-$n$ $[[n,1,n-1;n-1]]$ EA repetition codes do not saturate the quantum singleton bound or
 the linear programming bounds.
Were
an even-$n$ $[[n,1,n;n-1]]$ code to exist, it would have a weight enumerator $W_{\mathcal{L}\times \mathcal{S}_I}(x,y)=\sum_{w=0}^n B_w x^{n-w}y^{w}$ with $B_0=1, B_n=3,$ and $B_w=0$ for $w\neq 0,n$. The weight enumerator of its dual would also have
the coefficients
\[
A_w=\frac{1}{4}\left(3^w +3(-1)^w\right){n\choose w},
\]
which are positive integers for $w=0, \cdots, n$. It would only be able to
correct up to  $\left\lfloor {\frac{n-1}{2}}\right\rfloor $ channel qubit errors,
%the largest weight of correctable error operators for a code with minimum distance $n-1$ is $\lfloor \frac{n-2}{2}\rfloor$,
which is the same number of errors that our even-$n$ repetition codes can correct.
We prove below that even-$n$ $[[n,1,n;n-1]]$ EAQEC codes do not exist, and thus our even-$n$ repetition codes from Theorem~\ref{thm:even repetition} are optimal.
\begin{theorem}\label{thm:ne-rep}
There is no $\left[  \left[  n,1,n;n-1\right]  \right]  $ EAQEC code for $n$ even.
\end{theorem}
\begin{proof}
We prove this theorem by contradiction. Suppose there is an $\left[  \left[
n,1,n;n-1\right]  \right]  $ EAQEC code for $n$ even with a $2\times2n$
logical matrix%
\[
\left[  \left.
\begin{array}
[c]{c}%
u^{1}\\
u^{2}%
\end{array}
\right\vert
\begin{array}
[c]{c}%
v^{1}\\
v^{2}%
\end{array}
\right]  ,
\]
where $u^{1}$, $u^{2}$, $v^{1}$, and $v^{2}$ are binary row vectors of length $n$.
These vectors should satisfy the following condition in order for the above matrix to
be a valid logical matrix:%
\[
u^{1}\cdot v^{2}+u^{2}\cdot v^{1}=1\operatorname{mod}2.
\]
Let gw$\left(  \cdot\right)  $ be the \textquotedblleft general
weight\textquotedblright\ function defined by%
\[
\text{gw}\left(  u|v\right)  \equiv\sum_{i:u_{i}=1\text{ or }v_{i}=1}1,
\]
where $u_{i}$ denotes the $i^{\text{th}}$ bit of the binary $n$-tuple $u$. The
above binary vectors should satisfy the further constraints%
\[
\text{gw}\left(  u^{1}|v^{1}\right)  =\text{gw}\left(  u^{2}|v^{2}\right)
=\text{gw}\left(  u^{1}+u^{2}|v^{1}+v^{2}\right)  =n,
\]
in order for the code to have distance $n$ as claimed.

We now use the
above constraints  to obtain a contradiction. We first partition the
first row of the matrix into subsets $A$, $B$, and $C$ of $X$, $Y$, and $Z$
operators, respectively. There should not be any identity operators in the
first row in order for the code to have distance $n$. Up to permutations on
the qubits (under which the distance is invariant), the logical matrix has the
following form:%
\[
\left[  \left.
\begin{array}
[c]{ccc}%
\mathbf{1} & \mathbf{1} & \mathbf{0}\\
u_{A}^{2} & u_{B}^{2} & u_{C}^{2}%
\end{array}
\right\vert
\begin{array}
[c]{ccc}%
\mathbf{0} & \mathbf{1} & \mathbf{1}\\
v_{A}^{2} & v_{B}^{2} & v_{C}^{2}%
\end{array}
\right]  ,
\]
where $\mathbf{1}$ is a vector of all ones, $\mathbf{0}$ is a vector of all
zeros, and we have split up the vector $\left(  u^{2}|v^{2}\right)  $ into
different components corresponding to the subsets $A$, $B$, and $C$.
Consider the vector $u_{A}^{2}$.
Suppose that a component $\left(  u_{A}^{2}\right)_{i}=0$.
Then $\left(  v_{A}^{2}\right)  _{i}$ should equal $1$ so that the code's distance is not less than $n$.
Now suppose that $\left(  u_{A}^{2}\right)  _{i}=1$.
Then $\left(  v_{A}^{2}\right)  _{i}$ should also equal $1$ so that the code's distance is not less than $n$.
Otherwise, we could add $\left(  1|0\right)  $ to $(\left(  u_{A}^{2}\right)  _{i}|\left(  v_{A}^{2}\right)  _{i})$ and obtain $\left(  0|0\right)  $ as a component of
another logical operator, and such a result would imply that the code's
distance is less than $n$. These steps imply that the logical matrix should
have the following form:%
\[
\left[  \left.
\begin{array}
[c]{ccc}%
\mathbf{1} & \mathbf{1} & \mathbf{0}\\
u_{A}^{2} & u_{B}^{2} & u_{C}^{2}%
\end{array}
\right\vert
\begin{array}
[c]{ccc}%
\mathbf{0} & \mathbf{1} & \mathbf{1}\\
\mathbf{1} & v_{B}^{2} & v_{C}^{2}%
\end{array}
\right]  .
\]
Similar reasoning with $v_{C}^{2}$ and $u_{C}^{2}$ implies that $u_{C}^{2}$
should equal $\mathbf{1}$, and the logical matrix should then have the
following form:%
\[
\left[  \left.
\begin{array}
[c]{ccc}%
\mathbf{1} & \mathbf{1} & \mathbf{0}\\
u_{A}^{2} & u_{B}^{2} & \mathbf{1}%
\end{array}
\right\vert
\begin{array}
[c]{ccc}%
\mathbf{0} & \mathbf{1} & \mathbf{1}\\
\mathbf{1} & v_{B}^{2} & v_{C}^{2}%
\end{array}
\right]  .
\]
Finally, consider the vector $u_{B}^{2}$. Suppose that a component $\left(
u_{B}^{2}\right)  _{i}=1$. Then $\left(  v_{B}^{2}\right)  _{i}$ should equal
$0$ so that the code's distance is not less than $n$. Otherwise, we could add
$\left(  1|1\right)  $ to $(\left(  u_{B}^{2}\right)  _{i}|\left(  v_{B}%
^{2}\right)  _{i})$ and obtain $\left(  0|0\right)  $ as a component of
another logical operator, and such a result would imply that the code's
distance is less than $n$. Now suppose that $\left(  u_{B}^{2}\right)  _{i}%
=0$. Then $\left(  v_{B}^{2}\right)  _{i}$ should equal $1$, by reasoning
similar to the above. Thus, the logical matrix should have the following form
in order for the code's distance to be equal to $n$:%
\begin{equation}
\left[  \left.
\begin{array}
[c]{ccc}%
\mathbf{1} & \mathbf{1} & \mathbf{0}\\
u_{A}^{2} & u_{B}^{2} & \mathbf{1}%
\end{array}
\right\vert
\begin{array}
[c]{ccc}%
\mathbf{0} & \mathbf{1} & \mathbf{1}\\
\mathbf{1} & \bar{u}_{B}^{2} & v_{C}^{2}%
\end{array}
\right]  , \label{eq:ne-logical-matrix}
\end{equation}
where $\bar{u}_{B}^{2}$ is the binary complement of $u_{B}^{2}$. Now, the
symplectic product of the above two vectors is%
\[
\left(  \left\vert A\right\vert +\left\vert B\right\vert +\left\vert
C\right\vert \right)  \operatorname{mod}2=n\operatorname{mod}2=0,
\]
which contradicts the assumption that the original matrix is a valid logical matrix.
\end{proof}
\begin{theorem}\label{thm:ne-acc}
There is no $\left[  \left[  n,n-1,2;1\right]  \right]  $ EAQEC code for $n$ even.
\end{theorem}
\begin{proof}
We prove the theorem by contradiction, in a fashion similar to the previous theorem. Suppose there is an $[[n,n-1,2;1]]$
EAQEC code for $n$ even, and suppose its $2\times2n$ simplified check matrix
$[H_{X}|H_{Z}]$ has the form%
\[
\left[  \left.
\begin{array}
[c]{c}%
u^{1}\\
u^{2}%
\end{array}
\right\vert
\begin{array}
[c]{c}%
v^{1}\\
v^{2}%
\end{array}
\right]  ,
\]
where $u^{1}$, $u^{2}$, $v^{1}$, and $v^{2}$ are binary vectors of length $n$.
These vectors should satisfy the following condition in order for the above
matrix to be a simplified check matrix of a maximal-entanglement EAQEC\ code
with one ebit:%
\[
u^{1}\cdot v^{2}+u^{2}\cdot v^{1}=1\operatorname{mod}2.
\]
We now partition the first row of the simplified check matrix into subsets
$A$, $B$, and $C$ of $X$, $Y$, and $Z$ operators, respectively. Up to
permutations on the qubits (under which the distance is invariant), the
simplified check matrix has the following form:%
\[
\left[  \left.
\begin{array}
[c]{ccc}%
\mathbf{1} & \mathbf{1} & \mathbf{0}\\
u_{A}^{2} & u_{B}^{2} & u_{C}^{2}%
\end{array}
\right\vert
\begin{array}
[c]{ccc}%
\mathbf{0} & \mathbf{1} & \mathbf{1}\\
v_{A}^{2} & v_{B}^{2} & v_{C}^{2}%
\end{array}
\right]  ,
\]
where we have split up the vector $\left(  u^{2}|v^{2}\right)  $ into
different components corresponding to the subsets $A$, $B$, and $C$. The code
has minimum distance two by assumption, and is non-degenerate because it is a
maximal-entanglement EAQEC\ code. Therefore, no column of the above matrix
should be equal to the all-zeros vector. Were it not so, then the code would
not be able to detect every single-qubit $X$ or $Z$\ error and would not have
distance two as claimed. These constraints restrict the simplified check
matrix to have the following form:%
\[
\left[  \left.
\begin{array}
[c]{ccc}%
\mathbf{1} & \mathbf{1} & \mathbf{0}\\
u_{A}^{2} & u_{B}^{2} & \mathbf{1}%
\end{array}
\right\vert
\begin{array}
[c]{ccc}%
\mathbf{0} & \mathbf{1} & \mathbf{1}\\
\mathbf{1} & v_{B}^{2} & v_{C}^{2}%
\end{array}
\right]  .
\]
Also, no column of the entrywise sum of the matrices to the left and right of the
vertical bar should be equal to the all-zeros vector.
Were it not so, then the code would not be able to detect every single-qubit $Y$
error and would not have distance two as claimed. These constraints further
restrict the simplified check matrix to be as follows:%
\begin{equation}
\left[  \left.
\begin{array}
[c]{ccc}%
\mathbf{1} & \mathbf{1} & \mathbf{0}\\
u_{A}^{2} & u_{B}^{2} & \mathbf{1}%
\end{array}
\right\vert
\begin{array}
[c]{ccc}%
\mathbf{0} & \mathbf{1} & \mathbf{1}\\
\mathbf{1} & \bar{u}_{B}^{2} & v_{C}^{2}%
\end{array}
\right]  . \label{eq:check_matrix}
\end{equation}
Now, the symplectic product of the above two vectors is%
\[
\left(  \left\vert A\right\vert +\left\vert B\right\vert +\left\vert
C\right\vert \right)  \operatorname{mod}2=n\operatorname{mod}2=0,
\]
which contradicts the assumption that the original matrix is a simplified
check matrix for a maximal-entanglement EAQEC\ code with one ebit.
\end{proof}

The matching upper and lower bounds for $k=1$ in Table 1 of Ref. \cite{LBW13} are from the family of EA repetition codes.

Interestingly, observe that the non-existent logical matrix in (\ref{eq:ne-logical-matrix}) has the same form as the non-existent simplified check matrix in (\ref{eq:check_matrix}). Were either type of code to exist, we would expect them to be duals of each other, but they both fail to exist because they cannot satisfy the dual constraints imposed on them.

%Note that these codes are not the dual codes of the $n$-even repetition codes in Theorem \ref{thm:even repetition}.
%The second row of the simplified matrix in Theorem \ref{thm:distance_2code} anti-commutes
%with the row of the simplified check matrix corresponding to the last row of $H_1$ in Theorem \ref{thm:even repetition}.
%In addition,  the simplified logical matrix of an $n$-even repetition code has an all-zero column
%and consequently, the minimum distance of its dual code is 1.

\subsection{Existence of Other EAQEC Codes}

The following theorem is similar to Theorem 6 in Ref.~\cite{CRSS98}. It shows how to obtain
new EAQEC codes from existing ones. These results are helpful in our search for lower bounds
on the minimum distance of maximal-entanglement EAQEC codes.
\begin{theorem}  \label{thm:existence_nkdc}
Suppose an $[[n,k,d;c]]$ code exists. Then
\begin{enumerate}
    \item An $[[n+1,k,d;c+1]]$ code exists.

    \item An $[[n,k-1,d'\geq d;c+1]]$ code exists.

\end{enumerate}
\end{theorem}

\begin{proof}
\begin{enumerate}
    \item

Suppose $H=[H_X|H_Z]$ is a simplified check matrix of an $[[n,k,d;c]]$ code.
Then the simplified check matrix
\[
H'=\left[
\begin{array}{cc|cc}
00\cdots  0&0&00\cdots  0&1\\
00\cdots  0&1&00\cdots  0&0\\
 &0& &0\\
H_X &\vdots& H_Z&\vdots\\
 &0& &0\\
\end{array}
\right]
\]
defines an $[[n+1,k,d;c+1]]$ code.
We have the stabilizer group $(\mathcal{S}\otimes I) \cup \{ X_{n+1}, Z_{n+1}\}$,
where $(\mathcal{S}\otimes I)=\{E\otimes I: E\in \mathcal{S}\}$.
    \item It is obtained by moving a symplectic pair from the logical group to the stabilizer group.
\end{enumerate}

\end{proof}

%We can easily construct an $[[n,k,2;c]]$ EAQEC code, for example

\subsection{The Plotkin bound for EAQEC Codes}

The Plotkin bound for EAQEC codes is similar to the Plotkin bound for
classical codes \cite{MS77}. It is again helpful in our efforts to bound the
minimum distance of maximal-entanglement EAQEC codes. \begin{theorem} The Plotkin bound
for any $[[n,k,d;c]]$ EAQEC code is
\[
d\leq\frac{3n2^{2k-2}}{2^{2k}-1}.
\]
%\[
%d\leq\frac{3nM}{4(M-2^{n-k-c})},
%\]
%where $M=2^{n+k-c}$.

\end{theorem}
\begin{proof}
The proof is based on the proof of the classical Plotkin bound in Ref.~\cite{MS77}.
Let $M=2^{n+k-c}$ be the number of operators in $\mathcal{L}\times \mathcal{S}_I$. We bound the
quantity $\sum_{u,v\in\mathcal{L}\times \mathcal{S}_I\backslash \mathcal{S}_I}\mbox{wt}(u\cdot v)$ in two different ways.
First, we lower bound it. There are $M$ choices for $u$, and for each choice
of $u$, there are $M-2^{n-k-c}$ choices for $v$ such that $u\cdot v\notin \mathcal{S_I}$. Furthermore, for a code of minimum
distance $d$, $\mbox{wt}(u \cdot v)\geq d$ for any $u\cdot v\notin \mathcal{S_I}$. So the following lower
bound holds%
\begin{align*}
M(M-2^{n-k-c})d  & \leq\sum_{u,v\in\mathcal{L}\times \mathcal{S}_I:\ u\cdot v\notin \mathcal{S_I}}\mbox{wt}(u \cdot v)\\
& \leq  \sum_{u,v\in\mathcal{L}\times \mathcal{S}_I}\mbox{wt}(u \cdot v).
\end{align*}
The equality holds when $c=n-k$ because $S_I$ is trivial and wt$(u\cdot v) = 0$ if $u=v$.
Now we obtain an upper bound on the quantity. We form an $M\times n$ matrix
whose rows are the elements in the logical group $\mathcal{L}\times \mathcal{S}_I$. Let $m_{1}%
^{j},$ $m_{2}^{j},$ $m_{3}^{j}$, and $m_{4}^{j}$ be the number of $I$, $X$,
$Y$, and $Z$ operators in column $j$ of this matrix, respectively. So the
equality $\sum_{l=1}^{4}m_{l}^{j}=M$ holds for all $j\in\left\{
1,\cdots,n\right\}  $. Each choice of a particular Pauli operator and some
other Pauli operator in the same column contributes exactly 2 to the sum
$\sum_{u,v\in\mathcal{L}}\mbox{wt}(u\cdot v)$. Thus, the first equality below holds
for this reason, and the second holds by applying $\sum_{l=1}^{4}m_{l}^{j}=M$:%
\begin{align*}
\sum_{u,v\in\mathcal{L}\times \mathcal{S}_I}\mbox{wt}(u\cdot v) &  =\sum_{j=1}^{n}\sum_{l=1}^{4}%
m_{l}^{j}(M-m_{l}^{j})\\
&  =\sum_{j=1}^{n}\left(  M^{2}-\sum_{l=1}^{4}(m_{l}^{j})^{2}\right)  \\
&  \leq\sum_{j=1}^{n}\left(  M^{2}-\frac{M^{2}}{4}\right)  \\
&  =\frac{3n}{4}M^{2}.
\end{align*}
The first inequality follows by applying $\sum_{l=1}^{4}m_{l}^{j}/4=M/4$ and
convexity of the squaring function:$$\left(  M/4\right)  ^{2}=\left(
\sum_{l=1}^{4}m_{l}^{j}/4\right)  ^{2}\leq\sum_{l=1}^{4}\left(  m_{l}%
^{j}\right)  ^{2}/4 .$$
Combining the lower and upper bounds gives us the
EA Plotkin bound.
\end{proof}

Since the proof is independent of the number of ebits $c$, the EA Plotkin bound applies to arbitrary EAQEC codes.
However, note that $c$ does not appear in the bound, and consequently, this bound best describes the characteristics of maximal-entanglement EAQEC codes.
However, for large $k$, the bound is approximately $\frac{3}{4}n$.
Hence, this bound is useful only for small values of $k$.

\begin{remark} The EA Plotkin bound  has been improved in the case that an EAQEC code is ``linear" \cite{GL13_PhysRevA.87.032309}.
For EAQEC codes corresponding to classical linear quaternary codes, the linear EA Plotkin bound is
\[
d\leq \frac{3\cdot 2^{2k}}{8(2^{2k}-1)}(n+c+k).
\]
\end{remark}
%
%\subsection{Table of Lower and Upper Bounds on the Minimum Distance of Maximal-Entanglement EAQEC Codes}
%

%The upper bounds for $n\leq 15$ and $k\geq 2$ are from the linear programming bound,
%which is generally tighter than
%the Singleton bound:
% \[n-k+c\geq 2(d-1),\]
%and the Hamming bound for non-degenerate EAQEC codes \cite{Bowen02}:
%\[
%\sum_{j=0}^t 3^j {n \choose j} \leq 2^{n-k+c}.
%\]
The Plotkin bound and the linear programming bound match for $k\leq 2$ and $n\leq 15$.
For $k=3$ and $n=4,5,6,9,10,11,13,14,15$, they also match.
For $k>3$, the Plotkin bound is not as tight as the linear programming bound, the Singleton bound, or the Hamming bound.

\section{The Weight Enumerator Bound on the Block Error Probability
under Maximum A~Posteriori Decoding} \label{sec:error_bound}

Since maximal-entanglement codes bear many similarities to classical codes,
the block error probability when transmitting coded quantum information
through the depolarizing channel can be upper bounded using the weight
enumerator of a particular maximal-entanglement EAQEC code (similarly to the
case for classical codes \cite{RU08,McE02}). This \textquotedblleft weight
enumerator bound\textquotedblright\ gives an idea of the performance of
maximum-likelihood decoding of an arbitrary maximal-entanglement EAQEC code.
We can also determine the expected performance when decoding a random EAQEC
code with a maximum likelihood decoding rule. Below, we determine these bounds
and plot them for the maximal-entanglement repetition and accumulator
EAQEC\ codes. The result is that these codes perform comparably to a random
EA\ code with respect to this upper bound.

\begin{theorem}
\label{thm:particular-code-bound}
%Let $U$ be the Clifford encoder for the
%$\left[  \left[  n,k;n-k\right]  \right]  $ maximal-entanglement EAQEC\ code.
Suppose that a sender transmits an $\left[  \left[
n,k;n-k\right]  \right]  $ maximal-entanglement EAQEC\ code over a
depolarizing channel with parameter $p$, and furthermore, that the
receiver decodes this code according to a maximum a posteriori (MAP) decoding rule.
Then we have the following upper bound
on the block error probability $P_{B}$:%
\begin{align} \label{ieq:bound_BEP}
P_{B}\leq B\left(  \gamma\right)  -1,
\end{align}
where $B\left(  z\right) \triangleq W_{ \mathcal{L}}(1, z )  $ is the weight enumerator of the maximal-entanglement EAQEC\ code and $\gamma$ is the \textquotedblleft Bhattacharyya
parameter\textquotedblright\ for the depolarizing channel:%
\[
\gamma\equiv2\sqrt{\frac{p}{3}\left(  1-p\right)  }+\frac{2}{3}p.
\]
\end{theorem}

\begin{proof}
%Suppose that Alice wants to send a $k$-qubit state $|\phi\rangle$ to Bob and
%they share $(n-k)$ pairs of maximally-entangled states $|\Phi_{+}\rangle^{AB}=\frac{1}{\sqrt{2}}\left(  |00\rangle+|11\rangle\right)  $.
Let $U$ be a Clifford encoder for the $\left[  \left[  n,k;n-k\right]  \right]  $
maximal-entanglement EAQEC\ code. The encoded state $|\bar{\psi}\rangle^{AB}$
is
$
|\bar{\psi}\rangle^{AB}=(U^{A}\otimes I^{B})\left(  |\phi\rangle\otimes(|\Phi
_{+}\rangle^{AB})^{\otimes(n-k)}\right) .
$
%where the superscript $A$ or $B$ indicates that the operator acts on the
%qubits of Alice or Bob, respectively.
Then Alice transmits her qubits
(entangled with Bob's qubits) through $n$ independent uses of a depolarizing
channel $\mathcal{E}$ where
\[
\mathcal{E}\left(  \rho\right)  =\left(  1-p\right)  \rho+\frac{p}{3}\left(
X\rho X+Y\rho Y+Z\rho Z\right)  ,
\]
and $\rho$ is the density operator of a single qubit. We assume that $p<3/4$
because the channel is completely depolarizing when $p=3/4$.
Suppose that an
error operator $\tilde{E}\in\mathcal{G}^{n}$ occurs after the depolarizing
channel, and that $\mathbf{s^{x}},\mathbf{s^{z}}$ are the binary vector
representations of the error syndrome. Both $\mathbf{s^{x}}$ and $\mathbf{s^{z}}$ are of length $(n-k)$, and Bob
observes them by first decoding the qubits with a decoding unitary $U^{\dag}$ and
then performing Bell measurements on the ebits.
This implies that
\begin{align*}
(U^{\dag}\tilde{E})^{A}\otimes I^B |\bar{\psi}\rangle^{AB}
=&\tilde{L}_{0}|\phi\rangle\otimes\left(
\left(  X^{\mathbf{s^{x}}}Z^{\mathbf{s^{z}}}\right)  ^{A}\otimes
I^{B}\right) \left(  |\Phi_{+}\rangle^{AB}\right)  ^{\otimes n-k},
\end{align*}
and $\tilde{E}=U(\tilde{L}_{0}\otimes X^{\mathbf{s^{x}}}Z^{\mathbf{s^{z}}%
})U^{\dag}$ for some logical error $\tilde{L}_{0}\in\mathcal{L}_{0}$, where
$\mathcal{L}_{0}$ is the set of unencoded logical operators. Poulin \textit{et al}.~devised a {\it maximum a posteriori} decoder for standard stabilizer codes
\cite{PTO09}, \footnote{Poulin \textit{et al}. described their decoder as a ``maximum-likelihood" decoder \cite{PTO09},
% but a careful study of it reveals that their decoder is actually a maximum a posteriori decoder
but a careful study of it reveals that their decoder should more properly be called a maximum a posteriori decoder.}
and we can modify their decoder to be
%a maximal-entanglement EAQEC code maximum a posteriori decoder
a maximum a posteriori decoder $L_{\text{MAP}}\left(  \mathbf{s^{x}%
},\mathbf{s^{z}}\right)  $ for maximal-entanglement EAQEC codes, where%
\[
L_{\text{MAP}}\left(  \mathbf{s^{x}},\mathbf{s^{z}}\right)  \equiv
{\underset{L\in\mathcal{L}_{0}} {\arg\max}} \Pr\left\{  L|\mathbf{s^{x}},\mathbf{s^{z}}\right\}
.
\]
This decoder selects the most likely error operator acting on the logical qubits,
%recovery operation on the decoded qubits,
given the syndrome information $\mathbf{s^{x}}$ and $\mathbf{s^{z}}$. We can
calculate the above conditional distribution by applying the Bayes rule to the
joint distribution $\Pr\left\{  L,\mathbf{s^{x}},\mathbf{s^{z}}\right\}  $:%
\[
\Pr\left\{  L|\mathbf{s^{x}},\mathbf{s^{z}}\right\}  =\frac{\Pr\left\{
L,\mathbf{s^{x}},\mathbf{s^{z}}\right\}  }{\sum_{L^{\prime}}\Pr\left\{
L^{\prime},\mathbf{s^{x}},\mathbf{s^{z}}\right\}  },
\]
where%
\begin{align*}
\Pr\bigl\{L,  &  \mathbf{s^{x}},\mathbf{s^{z}}\bigr\}=\left.  \Pr\left\{
E\right\}  \right\vert _{E=U\left(  L\otimes X^{\mathbf{s^{x}}}%
Z^{\mathbf{s^{z}}}\right)  U^{\dagger}}\\
&  =\left.  \left(  1-p\right)  ^{n-\text{wt}\left(  E\right)  }\left(
\frac{p}{3}\right)  ^{\text{wt}\left(  E\right)  }\right\vert _{E=U\left(
L\otimes X^{\mathbf{s^{x}}}Z^{\mathbf{s^{z}}}\right)  U^{\dagger}}\\
&  =\left.  \left(  1-p\right)  ^{n}\left(  \frac{p}{3\left(  1-p\right)
}\right)  ^{\text{wt}\left(  E\right)  }\right\vert _{E=U\left(  L\otimes
X^{\mathbf{s^{x}}}Z^{\mathbf{s^{z}}}\right)  U^{\dagger}}.
\end{align*}
The distribution $\sum_{L^{\prime}}\Pr\left\{  L^{\prime},\mathbf{s^{x}},\mathbf{s^{z}}\right\}$ is fixed over all choices of $L$.
%satisfying the constraint $E=U\left( L\otimes X^{\mathbf{s^{x}}}Z^{\mathbf{s^{z}}}\right)  U^{\dagger}$ where $U$ is the encoding unitary.
Since $p<3/4\Longleftrightarrow p/\left(  3\left(
1-p\right)  \right)  <1$, the best choice of $L$ for the maximum a posteriori
decoder $L_{\text{MAP}}\left(  \mathbf{s^{x}},\mathbf{s^{z}}\right)  $ is the
one that selects a
%recovery operator $P$ with the minimum weight.
recovery operator $L^{-1}=L$ such that $E=U(L\otimes X^{\mathbf{s^{x}}%
}Z^{\mathbf{s^{z}}})U^{\dag}$ has the minimum weight. This minimum weight
decoder is similar to a classical minimum distance decoder.
% which is perhaps unsurprising because maximal-entanglement EAQEC codes bear many similarities
%to classical codes.

Let $\mathcal{L}$ be the set of encoded logical operators and $\mathcal{S}%
^{\prime}$ be the set of simplified stabilizer generators. Given $E_{0}\in\mathcal{L}_{0}$, let%
\[
Q(E_{0})\equiv\{\mathbf{s^{x}},\mathbf{s^{z}}:\Pr\{\tilde{L}%
_{0}E_{0},\mathbf{s^{x}},\mathbf{s^{z}}\}\geq\Pr\{\tilde{L}_{0},\mathbf{s^{x}%
},\mathbf{s^{z}}\}\}.
\]
We can now bound
the probability $P_{B}(\tilde{L}_{0})$ of a block error given that the error
operator $\tilde{L}_{0}$ occurs under this decoding scheme:%
\begin{align}
P_{B}(\tilde{L}_{0})= &  \Pr\{\mbox{MAP decoder fails}|\ \tilde{L}%
_{0}\mbox{ occurs}\}\nonumber\\
= &  \Pr\left\{  L_{\text{MAP}}\left(  \mathbf{s^{x}},\mathbf{s^{z}}\right)
\neq\tilde{L}_{0}\right\}  \nonumber\\
= &  \Pr\left\{  \tilde{L}_{0}\cdot L_{\text{MAP}}\left(  \mathbf{s^{x}%
},\mathbf{s^{z}}\right)  \neq I\right\}  \nonumber\\
= &  \Pr\left\{  \tilde{L}_{0}\cdot L_{\text{MAP}}\left(  \mathbf{s^{x}%
},\mathbf{s^{z}}\right)  \in\mathcal{L}_{0}\backslash I\right\}  \nonumber\\
= &  \sum_{E_{0}\in\mathcal{L}_{0}\backslash I}\Pr\left\{  \tilde{L}%
_{0}\cdot L_{\text{MAP}}\left(  \mathbf{s^{x}},\mathbf{s^{z}}\right)
=E_{0}\right\}  \nonumber\\
\leq & \sum_{E_{0}\in\mathcal{L}_{0}\backslash I}\sum_{\mathbf{s^{x}},\mathbf{s^{z}%
}\in Q(E_{0})}\Pr\left\{  \tilde{L}_{0},\mathbf{s^{x}%
},\mathbf{s^{z}}\right\}  .\nonumber
\end{align}%
Since $\sqrt{\frac{\Pr\{\tilde{L}_{0}E_{0},\mathbf{s^{x}},\mathbf{s^{z}}%
\}}{\Pr\{\tilde{L}_{0},\mathbf{s^{x}},\mathbf{s^{z}}\}}}\geq1$ for
$\mathbf{s^{x}},\mathbf{s^{z}}\in Q(E_{0})$, we can multiply
each term in sum by this factor and then
\begin{align*}
  P_{B}(\tilde{L}_{0})
&  \leq\sum_{E_{0}\in\mathcal{L}_{0}\backslash I}\sum_{\mathbf{s^{x}%
},\mathbf{s^{z}}\in Q(E_{0})}\sqrt{\Pr\left\{  \tilde{L}%
_{0},\mathbf{s^{x}},\mathbf{s^{z}}\right\}  \Pr\left\{  \tilde{L}_{0}%
E_{0},\mathbf{s^{x}},\mathbf{s^{z}}\right\}  }\\
&  \leq\sum_{E_{0}\in\mathcal{L}_{0}\backslash I}\sum_{\mathbf{s^{x}%
},\mathbf{s^{z}}\in\mathbb{Z}_{2}^{n-k}}\sqrt{\Pr\left\{  \tilde{L}%
_{0},\mathbf{s^{x}},\mathbf{s^{z}}\right\}  \Pr\left\{  \tilde{L}_{0}%
E_{0},\mathbf{s^{x}},\mathbf{s^{z}}\right\}  }\\
%\end{align*}
%\begin{align*}
&  =\sum_{E\in\mathcal{L}\backslash I}\sum_{M\in\mathcal{S}^{\prime}}\sqrt
{\Pr\left\{  \tilde{L}M\right\}  \Pr\left\{  \tilde{L}EM\right\}  }\\
&  \leq\sum_{E\in\mathcal{L}\backslash I}\sum_{M\in\mathcal{G}^{n}}\sqrt
{\Pr\left\{  \tilde{L}M\right\}  \Pr\left\{  \tilde{L}EM\right\}  },
\end{align*}
where $\tilde{L}=U(\tilde{L}_{0}\otimes I)U^{\dag}$, $E=UE_{0}U^{\dag}$, and
$M=U(I\otimes X^{\mathbf{s^{x}}}Z^{\mathbf{s^{z}}})U^{\dag}\in\mathcal{S}^{\prime}$.
Observe that
\begin{align*}
  \sum_{M\in\mathcal{G}^{n}}\sqrt{\Pr\left\{  \tilde{L}M\right\}  \Pr\left\{
\tilde{L}EM\right\}  }
= &  \sum_{M\in\mathcal{G}^{n}}\sqrt{\Pr\left\{  M\right\}  \Pr\left\{
\tilde{L}E\tilde{L}M\right\}  }\\
= &  \sum_{M\in\mathcal{G}^{n}}\prod_{i=1}^{n}\sqrt{\Pr\left\{  (M)_{i}%
\right\}  \Pr\left\{  (\tilde{L})_{i}(E)_{i}(\tilde{L})_{i}(M)_{i}\right\}
}\\
= &  \prod_{i=1}^{n}\sum_{(M)_{i}\in\mathcal{G}}\sqrt{\Pr\left\{
(M)_{i}\right\}  \Pr\left\{  (\tilde{L})_{i}(E)_{i}(\tilde{L})_{i}%
(M)_{i}\right\}  }.
\end{align*}
It holds that $(\tilde{L})_{i}(E)_{i}(\tilde{L})_{i}\neq I$ if $(E)_{i}\neq I$
and so%
\begin{align*}
&  \sum_{(M)_{i}\in\mathcal{G}}\sqrt{\Pr\left\{  (M)_{i}\right\}  \Pr\left\{
(\tilde{L})_{i}(E)_{i}(\tilde{L})_{i}(M)_{i}\right\}  }
  =2\sqrt{\frac{p}{3}(1-p)}+\frac{2}{3}p=\gamma.
\end{align*}
Otherwise,%
\[
\sum_{(M)_{i}\in\mathcal{G}}\sqrt{\Pr\left\{  (M)_{i}\right\}  \Pr\left\{
(\tilde{L})_{i}(E)_{i}(\tilde{L})_{i}(M)_{i}\right\}  }=1.
\]
Consequently,%
\begin{align*}
P_{B}(\tilde{L}_{0}) &  \leq\sum_{E\in\mathcal{L}\backslash I}\gamma
^{\text{wt}(\tilde{L}E\tilde{L})}\\
&  =\sum_{E\in\mathcal{L}\backslash I}\gamma^{\text{wt}(E)}\\
&  =B(\gamma)-1.
\end{align*}
Therefore, the probability $P_{B}$ of a block error is bounded by
$B(\gamma)-1$ when taking the expectation over all $\tilde{L}_{0}$.
\end{proof}

The above theorem is similar to Theorem~7.5 in Ref.~\cite{McE02}, which
determines an upper bound on the block error probability when transmitting a
classical linear code over a binary symmetric channel.

\begin{theorem} \label{thm:expected block error rate}
Suppose that the sender transmits
a random $\left[  \left[  n,k;n-k\right]
\right]  $ maximal-entanglement EAQEC\ code over a depolarizing channel with parameter $p$ and furthermore that
the receiver decodes this code according to a maximum a posteriori decoding
rule. Let $U$ be the Clifford encoder for this code.
Then we have the following upper bound on the expected block error
probability $\overline{P}_{B}$:%
\begin{equation}
\overline{P}_{B}=\mathbb{E}_{U}\left\{  P_{B}\right\}  \leq\frac{2^{2k}%
-1}{2^{2n}-1}\left(  \left(  1+3\gamma\right)  ^{n}-1\right)  ,\label{eq:WEB}%
\end{equation}
where $\gamma$ is  the  Bhattacharyya
parameter  defined in the previous theorem and the expectation is with
respect to the choice of random code. In particular, if the rate $k/n$
satisfies the following upper bound:%
\[
\frac{k}{n}<1-\frac{1}{2}\log_{2}\left(  1+3\gamma\right)  ,
\]
then the error probability decreases exponentially to zero in the asymptotic limit.
\end{theorem}

\begin{figure}
[ptb]
\begin{center}
\includegraphics[
natheight=4.373400in,
natwidth=5.833200in,
width=3.5405in
]%
{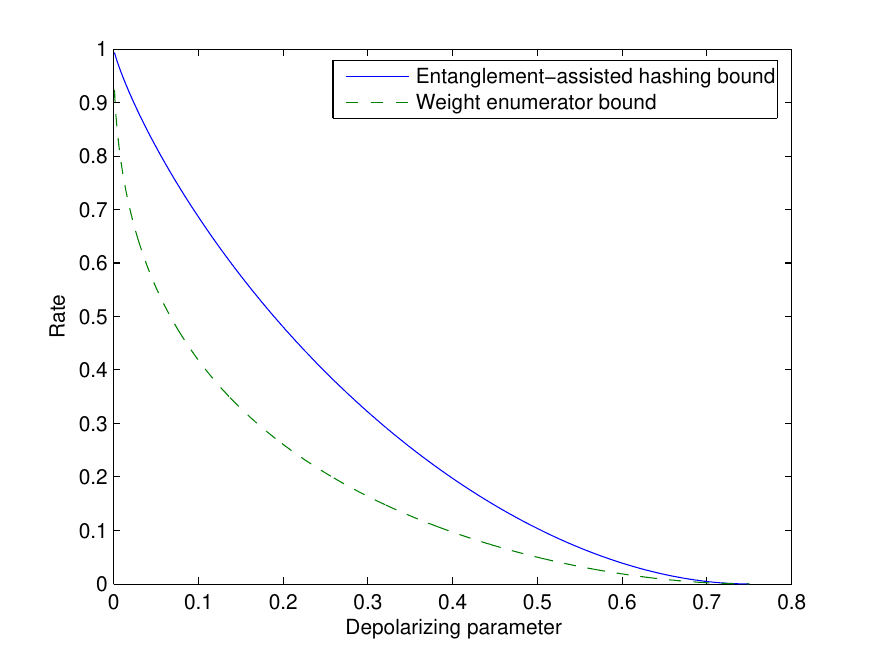}%
\caption{The figure plots both the EA hashing bound
$1-1/2\left[  H_{2}\left(  p\right)  +p\log_{2}3\right]  $ from
Ref.~\cite{PhysRevLett.83.3081} and the \textquotedblleft asymptotic weight
enumerator bound\textquotedblright\ from (\ref{eq:upper-MLD}) as a function of
the depolarizing parameter. The two bounds become close for high depolarizing
noise. Interestingly, the thresholds of the maximal-entanglement EA\ turbo
codes from Ref.~\cite{WH10}~are just shy of the asymptotic weight enumerator
bound (see Figures~6(b) and 7(b) of that paper).}%
\label{fig:EA-bounds}%
\end{center}
\end{figure}

\begin{proof}
We first establish a method for choosing a random maximal-entanglement EAQEC
code. A natural method for doing so is first to fix a basis of Pauli operators
$X_{1}$, $Z_{1}$, $X_{2}$, $Z_{2}$, \ldots, $X_{n}$, $Z_{n}$, where the first
$n-k$ anticommuting pairs correspond to the stabilizer operators for the $n-k$
ebits and the next $k$ anticommuting pairs correspond to the logical operators
for the $k$ information qubits. We then select a Clifford unitary uniformly at
random from the Clifford group (see Section~VI-A-2 of Ref.~\cite{DLT02} for a
relatively straightforward algorithm for doing so) and apply it to the above
fixed basis. This procedure produces $2n$ encoded operators $\overline{X}_{1}%
$, $\overline{Z}_{1}$, $\overline{X}_{2}$, $\overline{Z}_{2}$, \ldots,
$\overline{X}_{n}$, $\overline{Z}_{n}$ that specify a random
maximal-entanglement EAQEC code.

We now need to determine the expected weight enumerator $\mathbb{E}%
_{U}\left\{  B(z)\right\}  =\sum_{w=0}^{n}\mathbb{E}_{U}\left\{
B_{w}\right\}  z^{w}$ for  such a random maximal-entanglement code.
This will allow us to apply Theorem \ref{thm:particular-code-bound} to get an upper bound on the
expected block error probability. Each coefficient $\mathbb{E}_{U}\left\{
B_{w}\right\}  $ corresponds to the expected number of Pauli operators of
weight $w$ that belong to the logical operator group of a random EA\ code.
Equivalently, it corresponds to the expected number of Pauli operators of
weight $w$ that commute with the entanglement subgroup of a random code.
First, let us consider $\mathbb{E}_{U}\left\{  B_{0}\right\}  $. The identity
operator is the only Pauli operator with weight zero. It commutes with all
operators with unit probability. Thus, $\mathbb{E}_{U}\left\{  B_{0}\right\}
=1$. Now, let us consider $\mathbb{E}_{U}\left\{  B_{w}\right\}  $ with
$w\geq1$. We first determine the probability that a Pauli operator $g$ with
non-zero weight commutes with the $2\left(  n-k\right)  $ encoded operators
$\overline{X}_{1}$, $\overline{Z}_{1}$, $\overline{X}_{2}$, $\overline{Z}_{2}%
$, \ldots, $\overline{X}_{n-k}$, $\overline{Z}_{n-k}$ for a random EA\ code.
To simplify the calculation, observe that applying a uniformly random Clifford
unitary to the operators $X_{1}$, $Z_{1}$, $X_{2}$, $Z_{2}$, \ldots, $X_{n-k}%
$, $Z_{n-k}$ and then determining the probability that a fixed operator $g$
commutes with all of them is actually the same as keeping the basis fixed and
applying a random Clifford to the operator $g$ itself. This holds because%
%\[
%CfC^{\dag}g\overset{?}{\pm}gCfC^{\dag}=0\ \ \ \Longleftrightarrow
%\ \ \ fC^{\dag}gC\overset{?}{\pm}C^{\dag}gCf=0.
%\]
\[
CfC^{\dag}g{\pm}gCfC^{\dag}=0\ \ \ \Longleftrightarrow
\ \ \ fC^{\dag}gC{\pm}C^{\dag}gCf=0.
\]
Then a uniform distribution on the Clifford unitaries takes this operator $g$
to an arbitrary Pauli operator $g^{\prime}$, and the distribution induced is
just the uniform distribution on all of the $2^{2n}-1\ n$-qubit Pauli
operators not equal to the identity (this reasoning is the same as that in
Section~VI-A-1 of Ref.~\cite{DLT02}). At this point, the argument becomes
purely combinatorial, and the only operators that commute with the above fixed
basis are the ones with identity acting on the first $n-k$ qubits. Thus, there
are $2^{2k}-1$ Pauli operators besides the identity that commute with the
fixed basis, and we conclude that the probability that a fixed Pauli operator
$g$ with non-zero weight commutes with the random set $\overline{X}_{1}$, $\overline{Z}_{1}$,
$\overline{X}_{2}$, $\overline{Z}_{2}$, \ldots, $\overline{X}_{n-k}$,
$\overline{Z}_{n-k}$ is%
\[
\frac{2^{2k}-1}{2^{2n}-1}.
\]
Now we can calculate the expected number of operators that are in the logical
subgroup. The number of Pauli operators with weight $w$ is ${\binom{n}{w}%
}3^{w}$. Consequently, we have
\[
\mathbb{E}_{U}\{B_{w}\}=\frac{2^{2k}-1}{2^{2n}-1}\binom{n}{w}3^{w},
\]
which implies%
\begin{align*}
\mathbb{E}_{U}\left\{  B(z)\right\}  -\mathbb{E}_{U}\left\{  B_{0}\right\}
&  =\sum_{w=1}^{n}\mathbb{E}_{U}\left\{  B_{w}\right\}  z^{w}\\
&  =\frac{2^{2k}-1}{2^{2n}-1}\sum_{w=1}^{n}{\binom{n}{w}}3^{w}z^{w}\\
&  =\frac{2^{2k}-1}{2^{2n}-1}\left(  (1+3z)^{n}-1\right)  .
\end{align*}
Therefore, by exploiting the result in Theorem~\ref{thm:particular-code-bound}%
, an upper bound on the expected block error probability for general EAQEC
codes with maximal entanglement is%
\begin{align*}
\mathbb{E}_{U}\{P_{B}\} &  \leq\overline{B}(\gamma)-\overline{B}_{0},\\
&  =\frac{2^{2k}-1}{2^{2n}-1}\left(  \left(  1+3\gamma\right)  ^{n}-1\right)
.
\end{align*}
We can drive the expected error probability to be arbitrarily low in the large
$n$ and $k$ limit by ensuring that%
\begin{equation}
\frac{k}{n}<1-\frac{1}{2}\log_{2}\left(  1+3\gamma\right)
.\label{eq:upper-MLD}%
\end{equation}
This bound is not as tight as the EA hashing bound (the
optimal limit), and Figure~\ref{fig:EA-bounds}\ displays how these two bounds differ.
\end{proof}%

%TCIMACRO{\FRAME{ftbpFU}{3.5405in}{2.6602in}{0pt}{\Qcb{The figure plots both
%the EA hashing bound $1-1/2\left[  H_{2}\left(  p\right)
%+p\log_{2}3\right]  $ from Ref.~\cite{PhysRevLett.83.3081} and the
%\textquotedblleft asymptotic weight enumerator bound\textquotedblright\ from
%(\ref{eq:upper-MLD}) as a function of the depolarizing parameter. The two
%bounds become close for high depolarizing noise. Interestingly, the thresholds
%of the maximal-entanglement EA\ turbo codes from Ref.~\cite{WH10}~are just shy
%of the asymptotic weight enumerator bound (see Figures~6(b) and 7(b) of that
%paper).}}{\Qlb{fig:EA-bounds}}{ea-bounds.pdf}%
%{\special{ language "Scientific Word";  type "GRAPHIC";
%maintain-aspect-ratio TRUE;  display "USEDEF";  valid_file "F";
%width 3.5405in;  height 2.6602in;  depth 0pt;  original-width 5.8332in;
%original-height 4.3734in;  cropleft "0";  croptop "1";  cropright "1";
%cropbottom "0";  filename 'EA-bounds.pdf';file-properties "XNPEU";}}}%
%BeginExpansion
%EndExpansion

\begin{figure*}
[ptb]
\begin{center}
\includegraphics[
natheight=9.719600in,
natwidth=11.547000in,
width=6.5518in
]%
{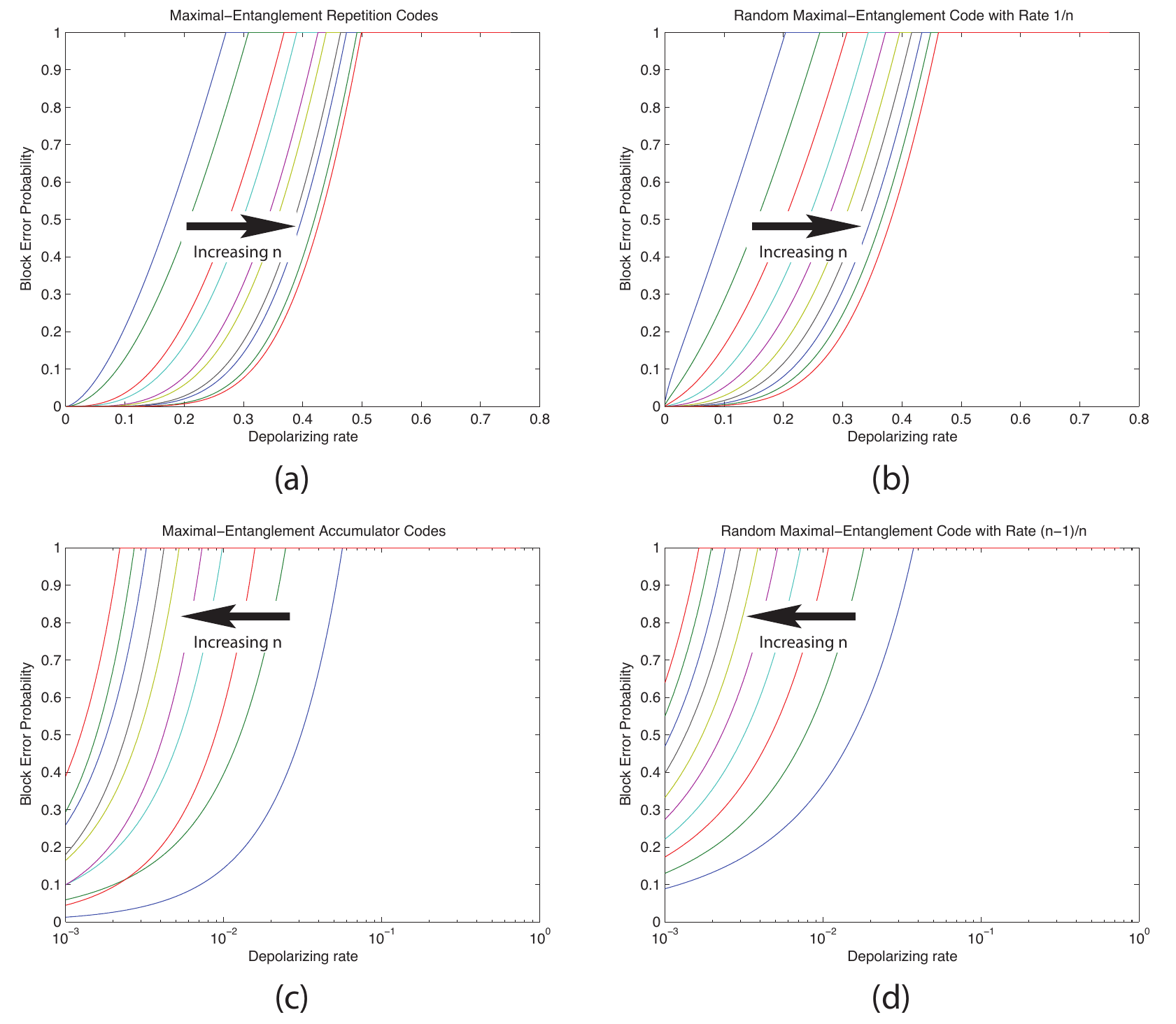}%
\caption{The figures plot the weight enumerator bound in (\ref{eq:WEB}) as a
function of the depolarizing parameter $p$ for various finite-length codes.
(a) The weight enumerator bound for maximal-entanglement repetition codes of
length 3 to 12. (b) The expected weight enumerator bound for random rate $1/n$
maximal-entanglement codes of length 3 to 12. (c) The weight enumerator bound
for maximal-entanglement accumulator codes of length 3 to 12. (d) The expected
weight enumerator bound for random rate $\left(  n-1\right)  /n$
maximal-entanglement codes of length 3 to 12. Observe that the performance of
the maximal-entanglement repetition and accumulator codes with respect to this
upper bound is comparable to the expected performance of random
maximal-entanglement codes.}%
\label{fig:error-bounds}%
\end{center}
\end{figure*}

We can plot the error probability bound in (\ref{ieq:bound_BEP}) as a function
of $p$ for specific codes such as the repetition codes or the accumulator
codes and then compare the results with the average error probability bound for a random code.
Figure \ref{fig:error-bounds} provides such plots and compares their performance with a random
EA\ code, with respect to these bounds.%
%TCIMACRO{\FRAME{ftbpFU}{6.5518in}{5.5201in}{0pt}{\Qcb{The figures plot the
%weight enumerator bound in (\ref{eq:WEB}) as a function of the depolarizing
%parameter $p$ for various finite-length codes. (a) The weight enumerator bound
%for maximal-entanglement repetition codes of length 3 to 12. (b) The expected
%weight enumerator bound for random rate $1/n$ maximal-entanglement codes of
%length 3 to 12. (c) The weight enumerator bound for maximal-entanglement
%accumulator codes of length 3 to 12. (d) The expected weight enumerator bound
%for random rate $\left(  n-1\right)  /n$ maximal-entanglement codes of length
%3 to 12. Observe that the performance of the maximal-entanglement repetition
%and accumulator codes with respect to this upper bound is comparable to the
%expected performance of random maximal-entanglement codes.}}%
%{\Qlb{fig:error-bounds}}{error-bounds.pdf}%
%{\special{ language "Scientific Word";  type "GRAPHIC";
%maintain-aspect-ratio TRUE;  display "USEDEF";  valid_file "F";
%width 6.5518in;  height 5.5201in;  depth 0pt;  original-width 11.547in;
%original-height 9.7196in;  cropleft "0";  croptop "1";  cropright "1";
%cropbottom "0";  filename 'error-bounds.pdf';file-properties "XNPEU";}}}%
%BeginExpansion
%EndExpansion

%\paragraph{Paragraph headings} Use paragraph headings as needed.

\section{Hashing Bounds for Pauli Channels}
\label{sec:hashing bounds}

The hashing bound of a quantum channel is an achievable rate for reliable
quantum communication \cite{BDSW96}, and as such,
it constitutes a lower bound on the quantum capacity
of a Pauli channel \cite{Lloyd97_PhysRevA.55.1613,Shor02,Devetak05}.
For a Pauli channel, this bound has a simple form and the proof that it is achievable is particularly simple as well.
In this section, we summarize several variations of the hashing protocol for
reliable communication in the asymptotic limit of many channel uses. In particular, one of the hashing bounds demonstrates that a maximal-entanglement EAQEC achieves the entanglement-assisted quantum capacity of a Pauli channel.
We consider  random stabilizer codes and use some techniques from the previous section.
For more details about quantum Shannon theory, we refer interested readers to Ref. \cite{Wilde11}
and references therein.

\subsection{Hashing Bound for Stabilizer Codes}
We first review a simple proof of the hashing bound for stabilizer codes \cite{Smith06}
 in order to have it available for helping to obtain the proofs of the hashing bounds for EAQEC codes.
\begin{theorem}
[Hashing Bound] There exists a quantum stabilizer code that
achieves the hashing limit $R=1-H\left(  \mathbf{p}\right)  $\ for a Pauli
channel  $\mathcal{E}$ of the following form:%
\[
\mathcal{E}(\rho)= p_{I}\rho+p_{X}X\rho X+p_{Y}Y\rho Y+p_{Z}Z\rho Z,
\]
where $\rho$ is the density operator of a single qubit, $\mathbf{p}=\left(  p_{I},p_{X},p_{Y},p_{Z}\right)  $ and $H\left(
\mathbf{p}\right)= -\sum_{i\in\{I,X,Y,Z\}}  p_{i} \log_2(p_i)  $ is the entropy of this probability vector.

\label{thm:hashing bound}
\end{theorem}

\begin{proof}
We need to correct only the typical errors. Define the
typical error set as follows:%
\[
T_{\delta}^{\mathbf{p}^{n}}\equiv\left\{  a^{n}:\left\vert -\frac{1}{n}%
\log_{2}\left(  \Pr\left\{  E_{a^{n}}\right\}  \right)  -H\left(
\mathbf{p}\right)  \right\vert \leq\delta\right\}  ,
\]
where $a^{n}$ is some sequence consisting of the letters $\left\{
I,X,Y,Z\right\}  $ and $\Pr\left\{  E_{a^{n}}\right\}  $ is the probability
that an  independent and identically distributed (IID) Pauli channel issues some tensor-product error $E_{a^{n}}\equiv
E_{a_{1}}\otimes\cdots\otimes E_{a_{n}}$, where $E_{a_j}\in \{I,X,Y,Z\}$.  This typical set consists of the
likely errors in the sense that%
\begin{equation}
\sum_{a^{n}\in T_{\delta}^{\mathbf{p}^{n}}}\Pr\left\{  E_{a^{n}}\right\}
\geq1-\epsilon,\label{eq:typical-errors}%
\end{equation}
for all $\epsilon>0$ and sufficiently large $n$. The quantum error correction
conditions for a stabilizer code defined by a stabilizer group $\mathcal{S}$ in this case are that $\{E_{a^{n}}:a^{n}\in
T_{\delta}^{\mathbf{p}^{n}}\}$ is a correctable set of errors if%
\[
E_{a^{n}}^{\dag}E_{b^{n}}\notin N\left(  \mathcal{S}\right)  \backslash
\tilde{\mathcal{S}},
\]
for all error pairs $E_{a^{n}}$ and $E_{b^{n}}$ such that $a^{n},b^{n}\in
T_{\delta}^{\mathbf{p}^{n}}$,
where $\tilde{\mathcal{S}}=\{eg: g\in \mathcal{S}, e\in\{\pm I, \pm iI\}  \}$. Also, we consider the expectation of the error
probability under a random choice of a stabilizer code and proceed to bound it as follows:%
\begin{align*}
\mathbb{E}_{\mathcal{S}}\left\{  p_{e}\right\}   &  =\mathbb{E}_{\mathcal{S}%
}\left\{  \sum_{a^{n}}\Pr\left\{  E_{a^{n}}\right\}  \mathcal{I}\left(
E_{a^{n}}\text{ is uncorrectable under }\mathcal{S}\right)  \right\}  \\
&  \leq\mathbb{E}_{\mathcal{S}}\left\{  \sum_{a^{n}\in T_{\delta}%
^{\mathbf{p}^{n}}}\Pr\left\{  E_{a^{n}}\right\}  \mathcal{I}\left(  E_{a^{n}%
}\text{ is uncorrectable under }\mathcal{S}\right)  \right\}  +\epsilon\\
&  =\sum_{a^{n}\in T_{\delta}^{\mathbf{p}^{n}}}\Pr\left\{  E_{a^{n}}\right\}
\mathbb{E}_{\mathcal{S}}\left\{  \mathcal{I}\left(  E_{a^{n}}\text{ is
uncorrectable under }\mathcal{S}\right)  \right\}  +\epsilon\\
&  =\sum_{a^{n}\in T_{\delta}^{\mathbf{p}^{n}}}\Pr\left\{  E_{a^{n}}\right\}
\Pr_{\mathcal{S}}\left\{  E_{a^{n}}\text{ is uncorrectable under }%
\mathcal{S}\right\}  +\epsilon.
\end{align*}
The first equality follows by definition: $\mathcal{I}$ is an indicator
function equal to one if $E_{a^{n}}$ is uncorrectable under~$\mathcal{S}$ and
equal to zero otherwise. The first inequality follows from
(\ref{eq:typical-errors}): we correct only the typical errors because the
atypical error set has negligible probability mass. The second equality
follows by exchanging the expectation and the sum. The third equality follows
because the expectation of an indicator function is the probability that the
event it selects occurs. Continuing, we have%
\begin{align*}
&  =\sum_{a^{n}\in T_{\delta}^{\mathbf{p}^{n}}}\Pr\left\{  E_{a^{n}}\right\}
\Pr_{\mathcal{S}}\left\{  \exists E_{b^{n}}:b^{n}\in T_{\delta}^{\mathbf{p}%
^{n}},\ b^{n}\neq a^{n},\ E_{a^{n}}^{\dag}E_{b^{n}}\in N\left(  \mathcal{S}%
\right)  \backslash\tilde{\mathcal{S}}\right\}  \\
&  \leq\sum_{a^{n}\in T_{\delta}^{A^{n}}}\Pr\left\{  E_{a^{n}}\right\}
\Pr_{\mathcal{S}}\left\{  \exists E_{b^{n}}:b^{n}\in T_{\delta}^{\mathbf{p}%
^{n}},\ b^{n}\neq a^{n},\ E_{a^{n}}^{\dag}E_{b^{n}}\in N\left(  \mathcal{S}%
\right)  \right\}  \\
&  =\sum_{a^{n}\in T_{\delta}^{\mathbf{p}^{n}}}\Pr\left\{  E_{a^{n}}\right\}
\Pr_{\mathcal{S}}\left\{  \bigcup\limits_{b^{n}\in T_{\delta}^{\mathbf{p}^{n}%
},\ b^{n}\neq a^{n}}E_{a^{n}}^{\dag}E_{b^{n}}\in N\left(  \mathcal{S}\right)
\right\}  \\
&  \leq\sum_{a^{n},b^{n}\in T_{\delta}^{\mathbf{p}^{n}},\ b^{n}\neq a^{n}}%
\Pr\left\{  E_{a^{n}}\right\}  \Pr_{\mathcal{S}}\left\{  E_{a^{n}}^{\dag
}E_{b^{n}}\in N\left(  \mathcal{S}\right)  \right\}  \\
&  \leq\sum_{a^{n},b^{n}\in T_{\delta}^{\mathbf{p}^{n}},\ b^{n}\neq a^{n}}%
\Pr\left\{  E_{a^{n}}\right\}  2^{-\left(  n-k\right)  }\\
&  \leq2^{2n\left[  H\left(  \mathbf{p}\right)  +\delta\right]  }2^{-n\left[
H\left(  \mathbf{p}\right)  +\delta\right]  }2^{-\left(  n-k\right)  }\\
&  =2^{-n\left[  1-H\left(  \mathbf{p}\right)  -k/n-\delta\right]  }.
\end{align*}
The first equality follows from the error correction conditions for a quantum
stabilizer code. %where $N\left(  \mathcal{S}\right)  $ is the normalizer of $\mathcal{S}$.
The first inequality follows by ignoring any potential
degeneracy in the code---we consider an error uncorrectable if it lies in the
normalizer $N\left(  \mathcal{S}\right)  $ and the probability can only be
larger because $N\left(  \mathcal{S}\right)  \backslash \tilde{\mathcal{S}}\subseteq N\left(
\mathcal{S}\right)  $. The second equality follows by realizing that the
probabilities for the existence criterion and the union of events are
equivalent. The second inequality follows by applying the union bound. The
third inequality follows from the fact that the probability for a fixed
operator $E_{a^{n}}^{\dag}E_{b^{n}}$ not equal to the identity commuting with
the stabilizer operators of a random stabilizer can be upper bounded as
follows:%
\[
\Pr_{\mathcal{S}}\left\{  E_{a^{n}}^{\dag}E_{b^{n}}\in N\left(  \mathcal{S}%
\right)  \right\}  =\frac{2^{n+k}-1}{2^{2n}-1}\leq2^{-\left(  n-k\right)  }.
\]
The reasoning here is similar to the reasoning in Theorem \ref{thm:expected block error rate}. %at the end of our other paper (arXiv:1010.5506).
 The random choice of a stabilizer code is equivalent to
fixing operators $Z_{1}$, \ldots, $Z_{n-k}$ and performing a uniformly random
Clifford unitary. The probability that a fixed operator commutes with
$\overline{Z}_{1}$, \ldots, $\overline{Z}_{n-k}$ is then just the number of
non-identity operators in the normalizer ($2^{n+k}-1$) divided by the total
number of non-identity operators ($2^{2n}-1$). After applying the above bound,
we then exploit the following typicality bounds:%
\begin{align*}
\forall a^{n}  & \in T_{\delta}^{\mathbf{p}^{n}}:\Pr\left\{  E_{a^{n}%
}\right\}  \leq2^{-n\left[  H\left(  \mathbf{p}\right)  +\delta\right]  },\\
\left\vert T_{\delta}^{\mathbf{p}^{n}}\right\vert  & \leq2^{n\left[  H\left(
\mathbf{p}\right)  +\delta\right]  }.
\end{align*}
We conclude that as long as the rate $k/n=1-H\left(  \mathbf{p}\right)
-2\delta$, the expectation of the error probability becomes arbitrarily small,
so that there exists at least one choice of a stabilizer code with the same
bound on the error probability.
\end{proof}

\subsection{Entanglement-Assisted Quantum Error-Correcting Codes}

\subsubsection{Maximal-Entanglement Codes}

Now consider the case of an EAQEC code. At first, we choose
the code to be a maximal-entanglement EAQEC code, so that
there are only information qubits or shares of ebits sent into the encoder.
The quantum error correction conditions in such a case become that $\{E_{a^{n}}\}$ is
a correctable set of errors if%
\[
E_{a^{n}}^{\dag}E_{b^{n}}\notin N\left(  \mathcal{S}_{S}\right)  ,
\]
for all error pairs $E_{a^{n}}$ and $E_{b^{n}}$ in the error set, where
$\mathcal{S}_{S}$ is the symplectic subgroup of the stabilizer code. It
follows for a random EAQEC code of this form that%
\[
\Pr_{\mathcal{S}}\left\{  E_{a^{n}}^{\dag}E_{b^{n}}\in N\left(  \mathcal{S}%
_{E}\right)  \right\}  =\frac{2^{2k}-1}{2^{2n}-1}\leq2^{-2\left(  n-k\right)
},
\]
because there are $2^{2k}-1$ nonidentity operators that commute with the $2\left(
n-k\right)  $ operators that generate $\mathcal{S}_{S}$. By modifying the last
few steps of the above proof as follows%
\begin{align*}
  \sum_{a^{n},b^{n}\in T_{\delta}^{\mathbf{p}^{n}},\ b^{n}\neq a^{n}}%
\Pr\left\{  E_{a^{n}}\right\}  2^{-2\left(  n-k\right)  }
 & \leq2^{2n\left[  H\left(  \mathbf{p}\right)  +\delta\right]  }2^{-n\left[
H\left(  \mathbf{p}\right)  +\delta\right]  }2^{-2\left(  n-k\right)  }\\
 & =2^{-2n\left[  1-H\left(  \mathbf{p}\right)  /2-k/n-\delta/2\right]  },
\end{align*}
we obtain the hashing bound for EAQEC codes:

\begin{theorem}
[EA Hashing Bound] There exists a maximal-entanglement EAQEC code that achieves the
EA hashing limit $R=1-H\left(  \mathbf{p}\right)  /2$\ for
a Pauli channel with parameters $\mathbf{p}$.
\end{theorem}

\subsubsection{Non-Maximal-Entanglement Codes}

We could also consider codes that do not use the maximal amount of ebits
possible. In this case, there are $k$ information qubits, $n-k-c$ ancilla
qubits, and $c$ ebits. The quantum error correction conditions in this case
become that $\{E_{a^{n}}\}$ is a correctable set of errors if%
\[
E_{a^{n}}^{\dag}E_{b^{n}}\notin N\left(  \mathcal{S}_{S},\mathcal{S}%
_{I}\right)  \backslash \tilde{\mathcal{S}}_{I},
\]
for all error pairs $E_{a^{n}}$ and $E_{b^{n}}$ in the error set, where
$\mathcal{S}_{S}$ is the symplectic  subgroup and $\mathcal{S}_{I}$ is the
isotropic subgroup of the EAQEC code, and $\tilde{\mathcal{S}}_I=\{eg: g\in \mathcal{S}_I, e\in\{\pm I, \pm iI\}  \}$. Focusing only on non-denegerate
errors, the error-correcting conditions become%
\[
E_{a^{n}}^{\dag}E_{b^{n}}\notin N\left(  \mathcal{S}_{S},\mathcal{S}%
_{I}\right)  .
\]
Then the relevant probability is%
\[
\Pr_{\mathcal{S}}\left\{  E_{a^{n}}^{\dag}E_{b^{n}}\in N\left(  \mathcal{S}%
_{S},\mathcal{S}_{I}\right)  \right\}  =\frac{2^{n+k-c}-1}{2^{2n}-1}%
\leq2^{-\left(  n-k+c\right)  }=2^{-n\left(  1-k/n+c/n\right)  },
\]
which follows from similar counting arguments.
%Let $Q=k/n$ and $E=c/n$.
This then leads to the following theorem for general EAQEC codes:
\begin{theorem}
[EA Hashing Region] There exists an EAQEC code whose achievable rate pair $\left(
Q=k/n,\right.$ $\left.E=c/n\right)  $ obeys the following EA hashing bound for a
Pauli channel with parameters $\mathbf{p}$:%
\begin{align*}
Q  & \leq1-H\left(  \mathbf{p}\right)  +E.
%Q  & \leq1-H\left(  \mathbf{p}\right)  /2.
\end{align*}
By varying $c$ from 0 to the maximal amount $n-k$, we can interpolate
between stabilizer codes and maximal-entanglement EAQEC codes and achieve
all rate pairs in the following hashing region:%
\begin{align*}
Q  & \leq1-H\left(  \mathbf{p}\right)+E,\\
Q  & \leq1-H\left(  \mathbf{p}\right)  /2.
\end{align*}

 \label{thm:hashing region}
\end{theorem}

\subsubsection{Entanglement-Assisted Codes with Imperfect Ebits}

In the case that the ebits of the receiver are not perfect, we
can use another stabilizer code to protect the ebits employed
in the EAQEC code for transmitting information qubits \cite{LB11a,WH10}.
Suppose that Alice uses an $[[n,k;c]]$ EAQEC code with a
(simplified) stabilizer group $\mathcal{S}_1$ through a Pauli
channel with parameter $\mathbf{p}_1$ to communicate with Bob and
Bob's qubits suffer a Pauli channel with parameter $\mathbf{p}_2$.
Furthermore, suppose Bob uses an $[[m,c]]$ stabilizer code with a
stabilizer group $\mathcal{S}_2$  to protect his $c$ qubits.
%Now we have two typical error sets
%\[
%T_{\delta}^{\mathbf{p_1}^{n}}\equiv\left\{  a^{n}:\left\vert -\frac{1}{n}%
%\log_{2}\left(  \Pr\left\{  E_{a^{n}}\right\}  \right)  -H\left(
%\mathbf{p_1}\right)  \right\vert \leq\delta\right\}  ,
%\]
%and
%\[
%T_{\delta}^{\mathbf{p_2}^{m}}\equiv\left\{  u^{m}:\left\vert -\frac{1}{m}%
%\log_{2}\left(  \Pr\left\{  E_{u^{m}}\right\}  \right)  -H\left(
%\mathbf{p_2}\right)  \right\vert \leq\delta\right\}  .
%\]

Suppose Bob uses two decoders in sequence to correct the errors---the first corrects the errors
on the ebits and the second corrects the errors on the information qubits.
Following the proof of Theorem~\ref{thm:hashing bound} and employing the union bound for two independent uses of the codes,
 we have the following hashing bound for combination
codes when the ebits are imperfect:
\begin{theorem}
[Hashing Bounds for Combination Codes] \label{thm:hashing bound combination}
Let $\alpha =\frac{m}{n}$. There exists an  $[[n,k;c]]$ EAQEC code combined with an $[[m,c]]$ stabilizer code
with achievable rate pair $\left(Q=k/n,\right.$ $\left.E=c/n\right)  $ obeys the following hashing bounds for
two Pauli channels with parameters $\mathbf{p}_1$ and $\mathbf{p}_2$, respectively:%
\begin{align*}
\frac{1}{\alpha}E&\leq 1- H(\mathbf{p_2}), \\
 Q & \leq 1- H(\mathbf{p_1})+E.
\end{align*}
\end{theorem}
On the other hand, Bob can treat the combination code as an
$[[n+m,k]]$ stabilizer code with a stabilizer group $\mathcal{S}$.
Using a similar argument as in the proof of Theorem~\ref{thm:hashing bound},
we find that $Q\leq 1+\alpha- H(\mathbf{p_1}) - \alpha H(\mathbf{p_2})$,
which agrees with Theorem \ref{thm:hashing bound combination} if the entanglement
consumption rate $E$ can be as large as $\alpha(1- H(\mathbf{p_2}))$.
This result might be considered surprising
because the simulations in Ref.~\cite{LB11a} suggest that a single decoder has
better performance than decoding the two codes
in sequence---however, it appears that this is a finite
blocklength effect that gets washed away in the asymptotic limit.

\subsection{EAQEC Codes for Classical Communication}

Now suppose the goal is to send classical data by exploiting
maximal-entanglement EAQEC codes. In this case, the
stabilizer structure is similar to that for a maximal-entanglement EAQEC code for
sending quantum data, but this time we do not care if $Z$ errors affect the
information qubits because they are classical. The error correction conditions
then become that $\{E_{a^{n}}\}$ is a correctable set of errors if%
\[
E_{a^{n}}^{\dag}E_{b^{n}}\notin N\left(  \mathcal{S}_{S},\mathcal{L}%
_{X}\right)  ,
\]
for all error pairs $E_{a^{n}}$ and $E_{b^{n}}$ in the error set, where
$\mathcal{S}_{S}$ is the symplectic subgroup and $\mathcal{L}_{X}$ is the
logical $X$ subgroup of the EAQEC code. Then the relevant probability is%
\[
\Pr_{\mathcal{S}}\left\{  E_{a^{n}}^{\dag}E_{b^{n}}\in N\left(  \mathcal{S}%
_{S},\mathcal{L}_{X}\right)  \right\}  =\frac{2^{k}-1}{2^{2n}-1}%
\leq2^{-\left(  2n-k\right)  }=2^{-n\left(  2-k/n\right)  },
\]
which follows from similar counting arguments. By modifying the last few steps
of the proof of Theorem~\ref{thm:hashing bound}, we obtain the following upper bound:%
\begin{align*}
  \sum_{a^{n},b^{n}\in T_{\delta}^{\mathbf{p}^{n}},\ b^{n}\neq a^{n}}%
\Pr\left\{  E_{a^{n}}\right\}  2^{-n\left(  2-k/n\right)  }
&  \leq2^{2n\left[  H\left(  \mathbf{p}\right)  +\delta\right]  }2^{-n\left[
H\left(  \mathbf{p}\right)  +\delta\right]  }2^{-n\left(  2-k/n\right)  }\\
&  =2^{-n\left[  2-H\left(  \mathbf{p}\right)  -k/n-\delta\right]  },
\end{align*}
giving the EA hashing bound for classical communication:

\begin{theorem}
[EA Hashing Bound for Classical Communication]There exists
an EAQEC code for classical
communication that achieves the EA classical hashing limit
$R=2-H\left(  \mathbf{p}\right)  $\ for a Pauli channel with parameters $\mathbf{p}$.
\end{theorem}

\section{Discussion} \label{sec:Discussion}

In this paper, we studied several properties of EAQEC codes, including the duality of EAQEC codes, the MacWilliams
identities for EAQEC codes, and the linear programming bound on the minimum distance of an EAQEC code. We
also derived the Plotkin bound and the Gilbert-Varshamov bound for
EAQEC codes, together with several theorems examining the
existence of EAQEC codes.
%Combining these results, we provided a table of upper and lower bounds on the minimum distance of
%maximal-entanglement EAQEC codes for $n\leq 15$.
Finally, we determined ``weight enumerator bounds'' on the block error
probability when decoding according to a maximum-likelihood
decoding rule, and we found that the performance of
maximal-entanglement repetition and accumulator codes is
comparable to the expected performance of random codes, with
respect to this upper bound.

The table of upper and lower bounds on the minimum distance of any
$[[n,k,d]]$ standard stabilizer codes ($c=0$) is given in
\cite{CRSS98}. Similar tables for EAQEC codes with $0<c<n-k$ can
be constructed.

We proposed a construction of $[[n,1,n-1;n-1]]$ EA repetition codes for $n$ even, which completes the family
of EA repetition codes for any $n$.
These EA repetition codes are the optimal codes that encode a single information qubit.
We also constructed an explicit encoding circuit for these codes.
%Most lower bounds in Table \ref{tb:Bounds}
%are from the optimization algorithm \cite{LB10}.
%However, these codes may not be close to optimal when $n+k$ becomes large,
% for the complexity of the encoding optimization algorithm increases exponentially with $n+k$, making full optimization impossible.
%To make the bounds in Table \ref{tb:Bounds} tighter, %we need to consider those entries whose upper bound and lower bound separate.
%we need to consider other code constructions to raise the lower bounds.
%One possibility is to  extend the encoding optimization algorithm from the form in Ref.~\cite{LB10}, by starting the optimization from a good EAQEC code and adding additional entanglement.
%From a given $[[n,k,d;c ]]$ code,
%we can change any subsets of the $n-k-c$ ancilla qubits with ebits and then apply a similar encoding optimization algorithm.
%From a given ``good'' $[[n,k,d_1;c ]]$ code, we obtain an optimized $[[n,k,d';n-k ]]$ code much faster than using the original optimization over an $[[n,k,d_2]]$ code.
%Since we begin with a ``good"  $[[n,k,d_1;c ]]$ code,  the optimized $[[n,k,d';n-k ]]$ code is expected to approach the upper bound. %on the minimum distance.
%Such a ``good" code can be obtained from the construction in Ref.~\cite{BDM06}, using optimal classical quaternary codes.
We also proved the non-existence of $[[n,1,n;n-1]]$ or $[[n,n-1,2;1]]$ codes for $n$ even,
which  decreases the upper bound predicted by the linear programming bound for $k=1$ and $n$ even.

We plan to explore the existence of other $[[n,k,d;n-k]]$ codes to decrease the upper bound.
%Given a $n\times m$ quaternary matrix, we can use this as a simplified check matrix of an $[[n,n-m+c,d;c]]$ EAQEC code.
%provided that the number of symplectic pairs $c$ satisfies  that \[c\leq \frac{m}{2}.\]
%We cannot follow the proof of classical Gilbert-Varshamov bound  to construct a simplified check matrix because $c$.
%One might try to prove a  G-V bound by considering the Hamming ball of the quantum states.
%Shadow enumerators, designed for self-dual codes, give additional constraints on the weight
%enumerator of a stabilizer code. However, there is no general form in the case of EAQEC codes.
Consider the  possibility of a ``self-dual code'' $[[n, n/2,d; n/2]]$ for $n$
even, such that the dual code is also an $[[n,n/2,d; n/2]]$ code with the same
weight enumerators. That is, $W_{ \mathcal{L}}(x,y)=W_{ \mathcal{S}_S}(x,y)$.
We conjecture that such self-dual codes exist. % with the minimum distance $d= \frac{n}{2}$.
If so, the two groups $\mathcal{S}_S$ and $\mathcal{L}$ may be equivalent up to a permutation on the qubits.
 Such codes would have interesting and useful properties.

Finally, we applied the idea of random stabilizer codes to prove an upper bound on the average block error rate, and
we also proved several variations of the hashing bound for EAQEC codes.
It should be possible to improve upon the hashing regions by exploiting degeneracy in
EAQEC codes, by an approach similar to that from Ref.~\cite{SS07_PhysRevLett.98.030501}.

%ADD\ IN\ LATER ABOUT\ HASHING\ BOUND\ FOR\ COMBINATION\ CODES...

%\begin{acknowledgements}
TAB and CYL were supported in part by NSF Grant CCF-0830801.
This work was supported in part by the Intelligence Advanced Research Projects Activity (IARPA) via Department of Interior National Business Center contract number D11PC20165. The U.S. Government is authorized to reproduce and distribute reprints for Governmental purposes notwithstanding any copyright annotation thereon. The views and conclusions contained herein are those of the authors and should not be interpreted as necessarily representing the official policies or endorsements, either expressed or implied, of IARPA, DoI/NBC or the U.S. Government.
MMW\ acknowledges the support of the MDEIE (Qu\'{e}bec) PSR-SIIRI international collaboration grant.
MMW acknowledges useful discussions with Omar Fawzi and Jan Florjanczyk.
%\end{acknowledgements}

%% BibTeX users please use one of
%%\bibliographystyle{spbasic}      % basic style, author-year citations
%\bibliographystyle{spmpsci}      % mathematics and physical sciences
%%\bibliographystyle{spphys}       % APS-like style for physics
%\bibliography{../qecc}   % name your BibTeX data base

\end{document}